# To See a World in a Grain of Sand

—— The Scientific Life of Shoucheng Zhang

*Biao Lian, Chao-Xing Liu, Xiao-Qi Sun, Steven Kivelson, Eugene Demler and Xiao-Liang Qi*

Our friend and colleague, Prof. Shoucheng Zhang, passed away in 2018, which was a great loss for the entire physics community. For all of us who knew Shoucheng, it is difficult to overcome the sadness and shock of his early departure. However, we are very fortunate that Shoucheng has left us such a rich legacy and so many memories in his 55 years of life as a valuable friend, a world-leading physicist, a remarkable advisor, and a great thinker. On May 2-4, 2019, a memorial workshop for Shoucheng was organized at Stanford University, where we displayed a small exhibition of 12 posters, as a brief overview of Shoucheng's wonderful scientific life. This article is prepared based on those posters.

## 1. Early Experience

### 1.1. *Childhood and school age*

Shoucheng Zhang (张首晟) was born in 1963 in Shanghai, China. In his childhood, Shoucheng already showed exceptional talent and a strong interest in various fields of knowledge. In the attic of Shoucheng's home there were many books on art, history, philosophy, and science left by his grandfather and others of his parent's generation. These included books on the philosophy of Russell and Kant and the art of da Vinci and Rodin. Shoucheng's favorite activity after school was to read books in the attic. In an era when educational resources were scarce, these books opened a new world to him. In 1976, Shoucheng's father bought Shoucheng a set of high school textbooks on mathematics, physics, and chemistry. He was immediately attracted by the amazing beauty of science.

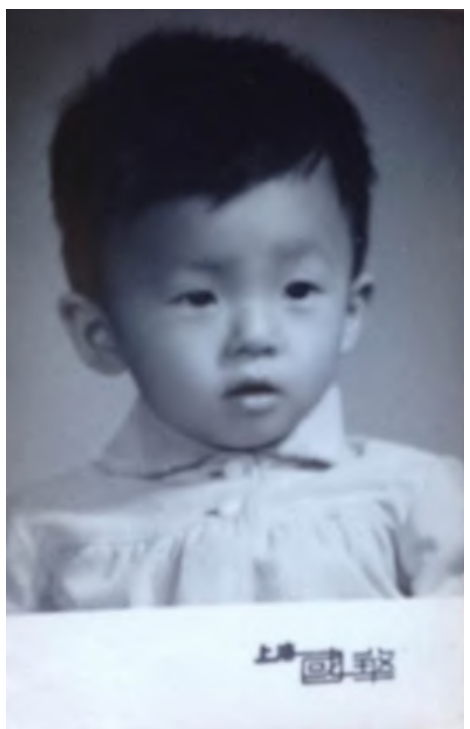

**Figure 1. Shoucheng at age two.**

### 1.2. *College time in Shanghai and Berlin*

In 1977, the National Higher Education Entrance Exam of China was restarted after the end of the Cultural Revolution. Without attending high school, Shoucheng took the first exam and got admitted to the Physics Department of Fudan University in Shanghai in 1978. At the age of 15, he was the youngest student in his class. One year later, in recognition of his excellent academic performance, Shoucheng was selected for an exchange program to study abroad at the Free University of Berlin, where he received his *Diplom-Physiker* (Bachelor of Science degree) in 1983.

During his college time in Germany, besides studying physics, Shoucheng also had rich exposure to German culture. On a trip back from Bonn to Berlin in 1981, he and some friends visited the Stadtfriedhof cemetery in Göttingen that houses the graves of many scientists. Even after many years, Shoucheng often remembered this visit as a source of inspiration. Upon observing the equations on the tombstones of Max Planck, Otto Hahn, and Max Born, Shoucheng wrote that he would "spend my energy on the pursuit of science, hoping that I too would leave behind a life's work that could be summed up in a simple equation."

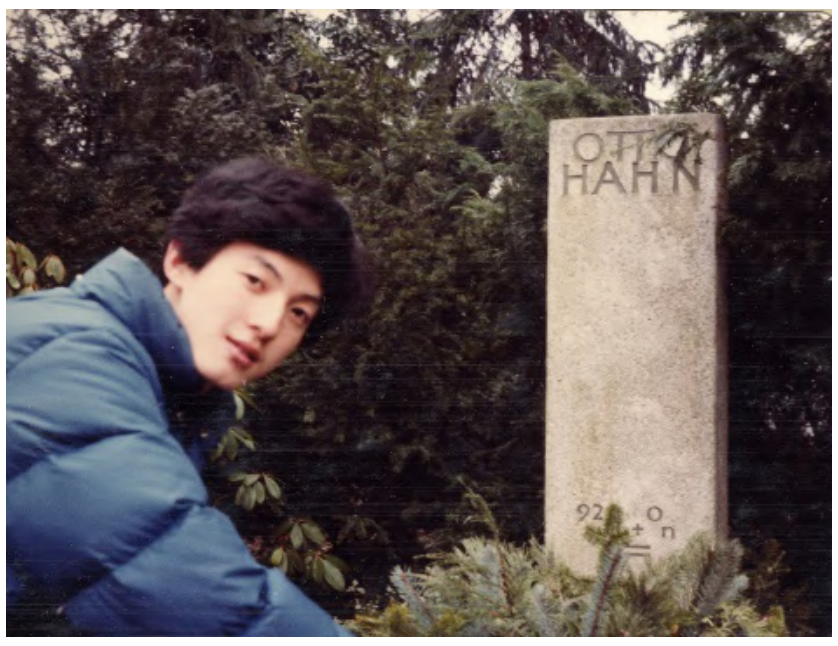

**Figure 2. Shoucheng in front of the grave of nuclear physicist Otto Hahn in Göttingen, 1981.**

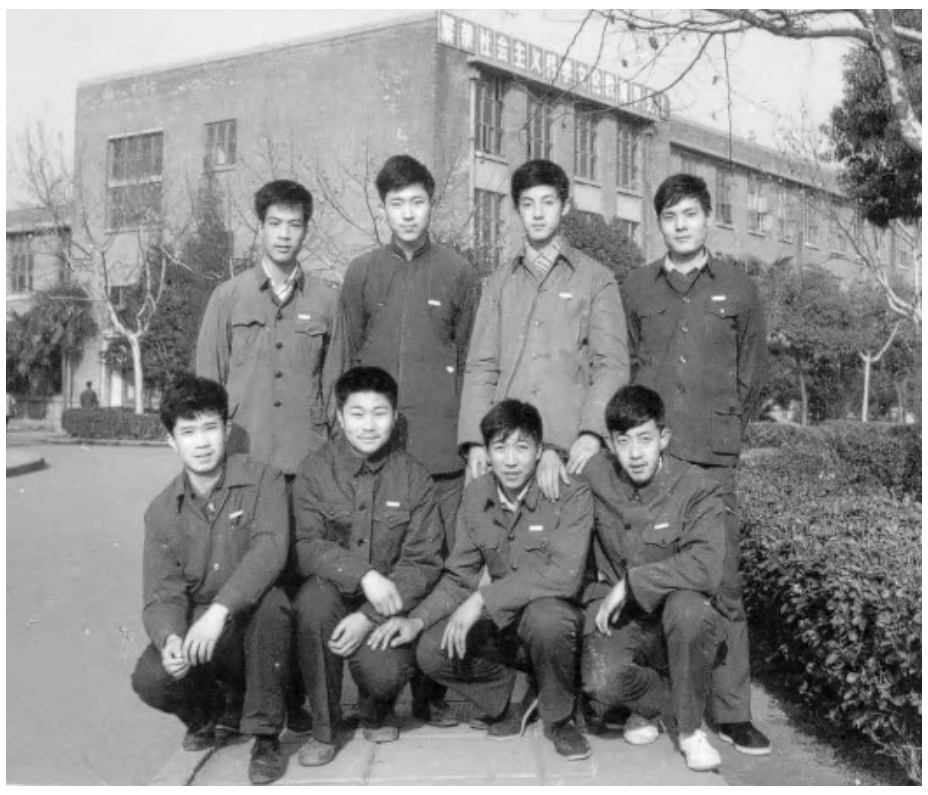

**Figure 3 Shoucheng (second from the right in the second row) and his roommates at Fudan University, 1978**

## 1.3. *Ph.D. at Stony Brook University*

Shoucheng began his Ph.D. studies on supergravity at the State University of New York at Stony Brook in 1983, advised by Peter van Nieuwenhuizen. In the final year of Shoucheng's Ph.D. (1986-1987), following Prof. Chen-Ning Yang's suggestion, he started shifting his research direction to condensed matter physics. He began a collaboration with Steven Kivelson, a faculty member at Stony Brook who later became Shoucheng's colleague and lifelong friend at Stanford.

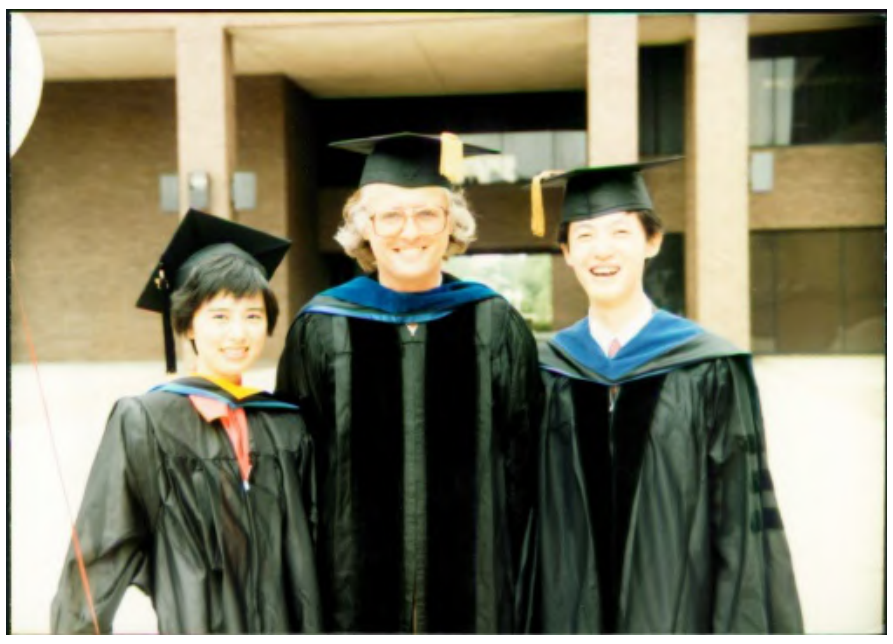

**Figure 5 Shoucheng with his advisor Peter van Nieuwenhuizen (middle) and his future wife Barbara (left) at their graduation ceremony, 1987.**

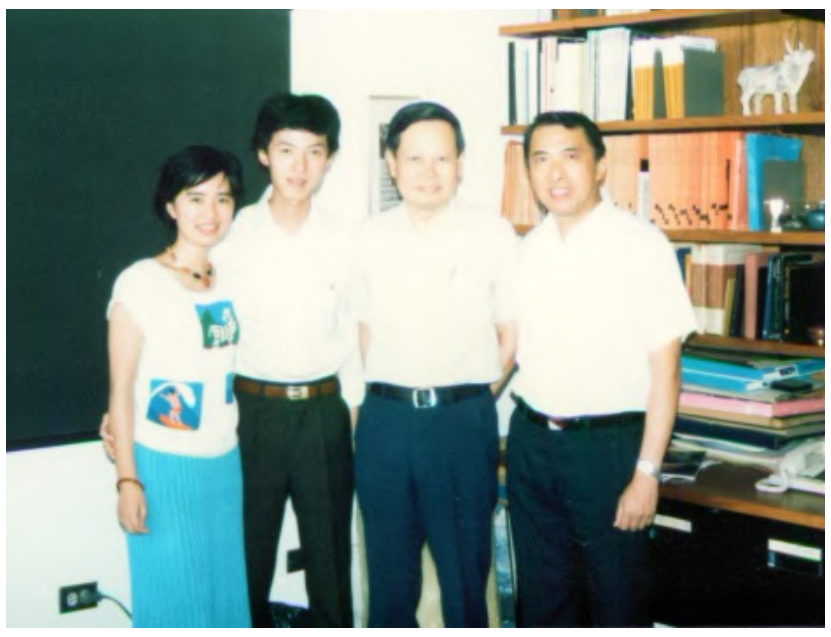

**Figure 4 Barbara, Shoucheng, Chen-Ning Yang, and Shoucheng's father Hongfan, 1987**

## 1.4. *Santa Barbara, IBM and Stanford*

After receiving his Ph.D. in 1987 from Stony Brook, Shoucheng became a postdoctoral fellow at the Institute of Theoretical Physics (ITP) in UC Santa Barbara. In 1987, he married his childhood sweetheart Barbara Yu. He then joined IBM Almaden Research Center as a Research Staff Member from 1989 to 1993. Thereafter, he joined the faculty of Stanford University, which remained his academic home for the rest of his life.

During his career, Shoucheng has worked on many different directions in physics, such as the theory of fractional quantum Hall states and high temperature superconductivity, topological insulators and topological superconductors, spintronics, etc. In Table 1, we provide a sketch of some key events in his life and career Hisscientific achievements in different areas of physics will be overviewed in the rest of this article.

**Table 1: Some key events in Shoucheng's career and life.**

| Year | Events |
|---|---|
| 1963 | Shoucheng was born in Shanghai, China to Manfan Ding and Hongfan Zhang |
| 1980 | Began studying at Free University of Berlin |
| 1983 | Graduated from Free University of Berlin<br>Started Ph.D. at Stony Brook University |
| 1986 | Started working on condensed matter physics |
| 1987 | Received Ph.D. and started postdoc at ITP, UC Santa Barbara<br>Married his childhood sweetheart, Barbara Yu<br>Chern-Simons theory of fractional quantum Hall states |
| 1989 | Started at IBM Almaden Research Center, San Jose<br>Global phase diagram of fractional quantum Hall states |
| 1993 | Joined the faculty of Stanford University<br>Shoucheng's son Brian was born |
| 1996 | Shoucheng's daughter Stephanie was born |
| 1997 | Proposed the SO(5) theory of high Tc superconductors |
| 2001 | Generalized quantum Hall effect to 4 dimensions |
| 2003 | Proposed the intrinsic spin Hall effect |
| 2004-2005 | Early models of quantum spin Hall effect (2004-2005) |
| 2006 | Predicted the quantum spin Hall effect in HgTe |
| 2007 | Quantum spin Hall effect realized in HgTe |
| 2008-2009 | Topological magneto-electric effect<br>Prediction and realization of $Bi_2Te_3$ family of topological insulators<br>Prediction of new topological superconductors |
| 2010 | Prediction of quantum anomalous Hall effect |
| 2013 | Quantum anomalous Hall effect realized<br>Founded venture capital firm DHVC<br>Celebrated 50th birthday with his friends, current and past group members. |
| 2017 | Experimental evidence of chiral topological superconductor reported |
| 2018 | Founded the Stanford Center for Topological Quantum Physics<br>Passed away on December 1, 2018 |

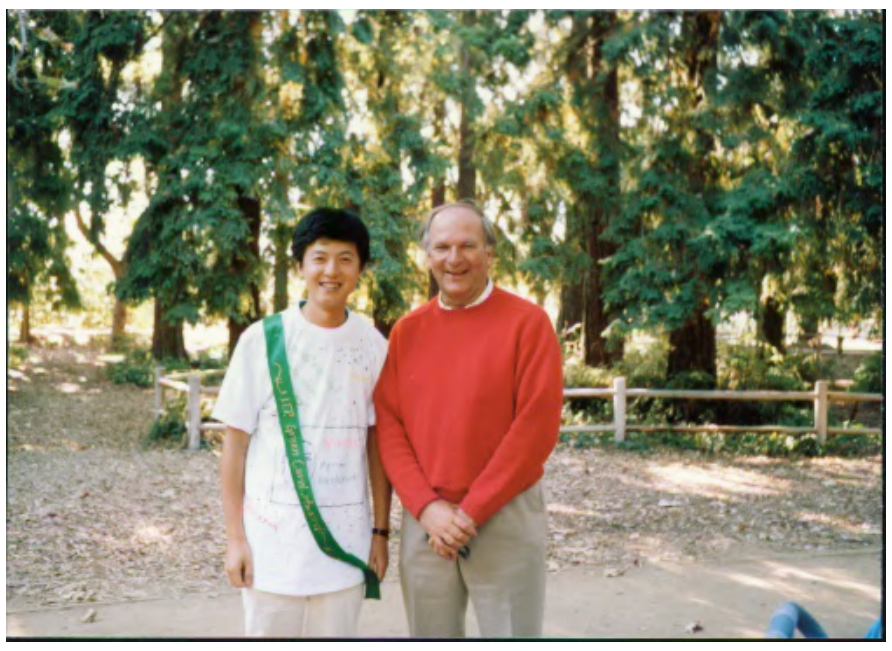

**Figure 7 Shoucheng and J. Robert Schrieffer at Santa Barbara, 1989**

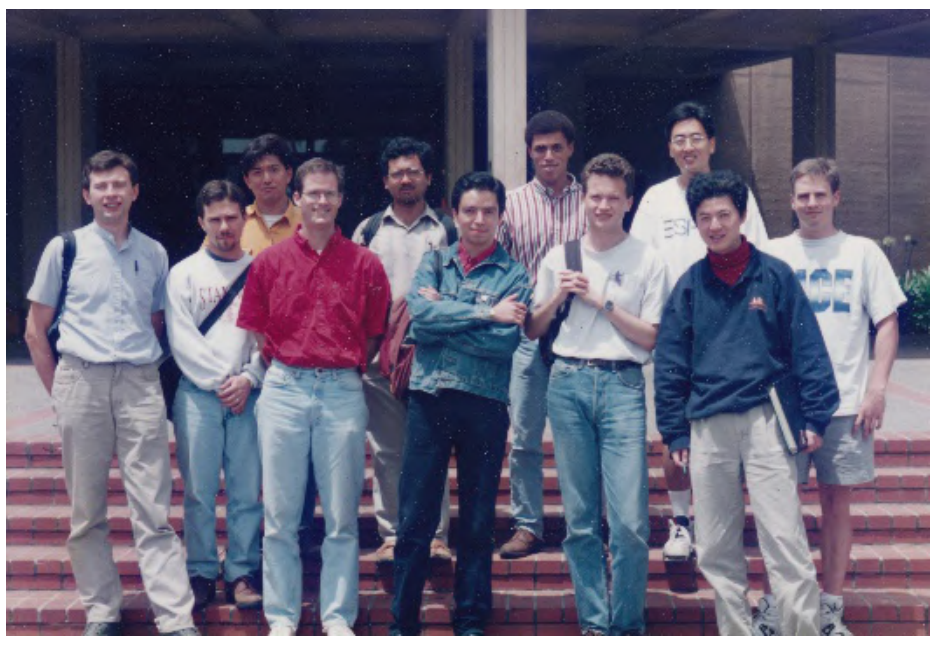

**Figure 6 A group photo of Shoucheng and his students after a class at Stanford in 1994.**

## 2. Theory of Fractional Quantum Hall Effect

### 2.1. *The fractional quantum Hall effect*

When Shoucheng began to study condensed matter physics in the late 1980s, one of the first topics he became interested in was the fractional quantum Hall effect (FQHE).

The FQHE, experimentally discovered in 1982 by D. C. Tsui, H. L. Stormer, and A. C. Gossard [41], is a remarkable quantum phenomenon of two dimensional (2D) metals near the absolute zero of temperature in a strong magnetic field $B$, where the Hall resistance $R_{xy}$ is quantized at value

$$R_{xy} = h/\nu e^2$$

Here $\nu$ is one of a particular set of rational fractions, $h$ is Planck's constant, and $e$ is the electron charge.

The understanding of the FQHE fundamentally challenged conventional condensed matter theory. In 1983, Robert B. Laughlin [43] proposed the Laughlin wavefunction which successfully described the ground state of fractional quantum Hall fluids with $\nu = \frac{1}{2k-1}$ ( $k = 1,2,\cdots$ ); this laid the foundation for the theory of the FQHE.

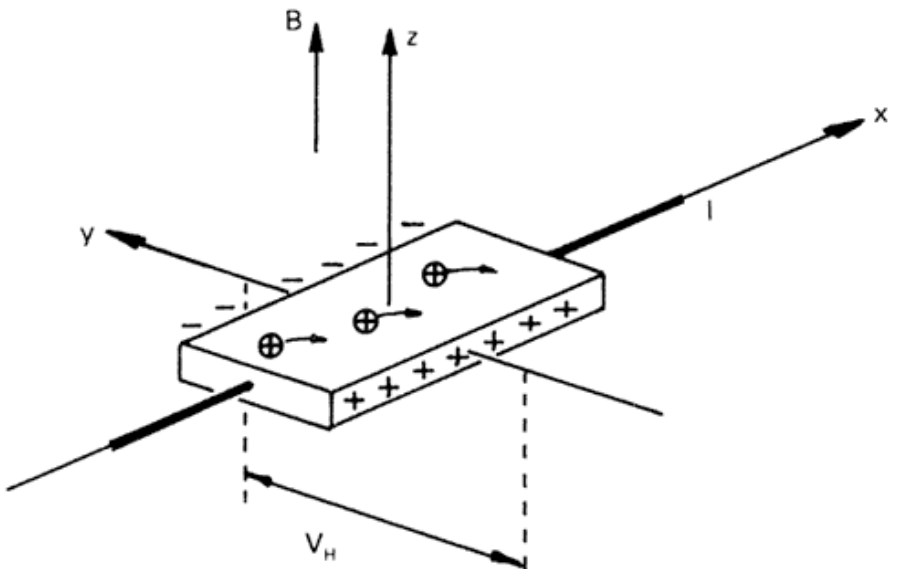

**Figure 9 The experiment of Hall resistance $R_{xy} = V_y/I_x$ (Kosmos 1986).**

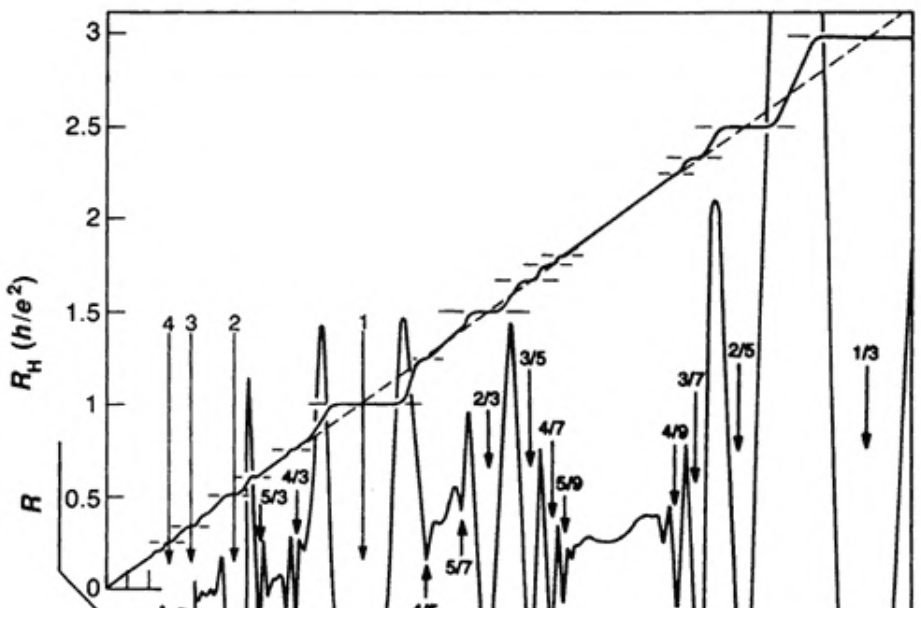

**Figure 8 The FQHE with $R_{xy} = h/\nu e^2$, from Ref. [10]**

### 2.2. *Chern-Simons Ginzburg-Landau Theory and the Global phase diagram*

Shoucheng was interested in looking for a quantum field theory (QFT) for the FQHE. The presence of the background magnetic field $B$ implies that this field theory must be quite different from those that were familiar in high energy physics. The field theory needed to capture essential features of the fractional quantum Hall state such as the quantized Hall conductance, fractional charged excitations, etc.
In 1989, Shoucheng and his collaborators Thors Hans Hansson and Steven Kivelson [44] proposed the Chern-Simons-Landau-Ginzburg field theory of the FQHE:

$$\mathcal{L} = \mathcal{L}_{GL}\left[\phi, \left(i\partial_\mu - A_\mu + a_\mu\right)\phi\right] + \left(\frac{e\pi}{2\theta\Phi_0}\right)\epsilon^{\mu\nu\lambda}a_\mu\partial_\nu a_\lambda \,.$$

The key idea of this theory is **flux attachment**, which is achieved by introducing a gauge field with a Chern-Simons term. The dynamics of this gauge field attaches $2k-1$ magnetic fluxes to each electron, which transmutes them into bosons. The FQHE is then interpreted as the condensate of this boson.

In an interview with Stanford News after Shoucheng's death, Steven Kivelson described this discovery: "One day Shoucheng came to visit and he said, 'Look what I figured out.' He then sketched out a basic idea of the theory and all of these mysterious features of the fractional quantum Hall effect just dropped in your lap incredibly simply," Kivelson said. "That's not how physics usually works. You usually slave away at things. But on that day Shoucheng's idea was just so focused and so perfect."

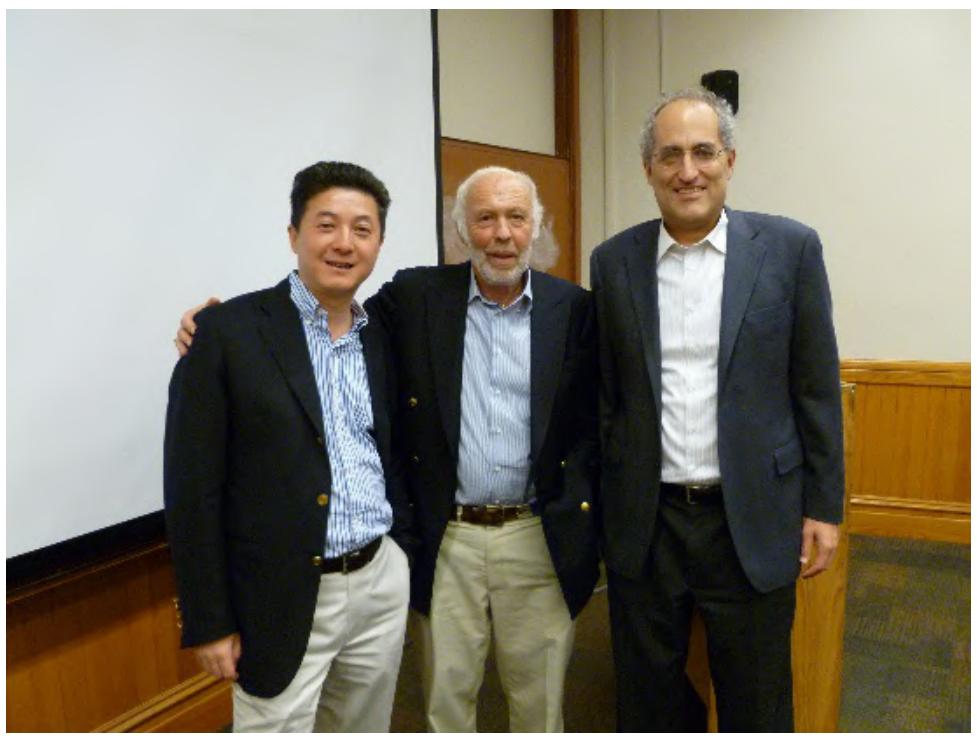

**Figure 10 Shoucheng with James H. Simons (middle), one of the founders of Chern-Simons theory, and Edward Witten (right) at Stanford in 2010.**

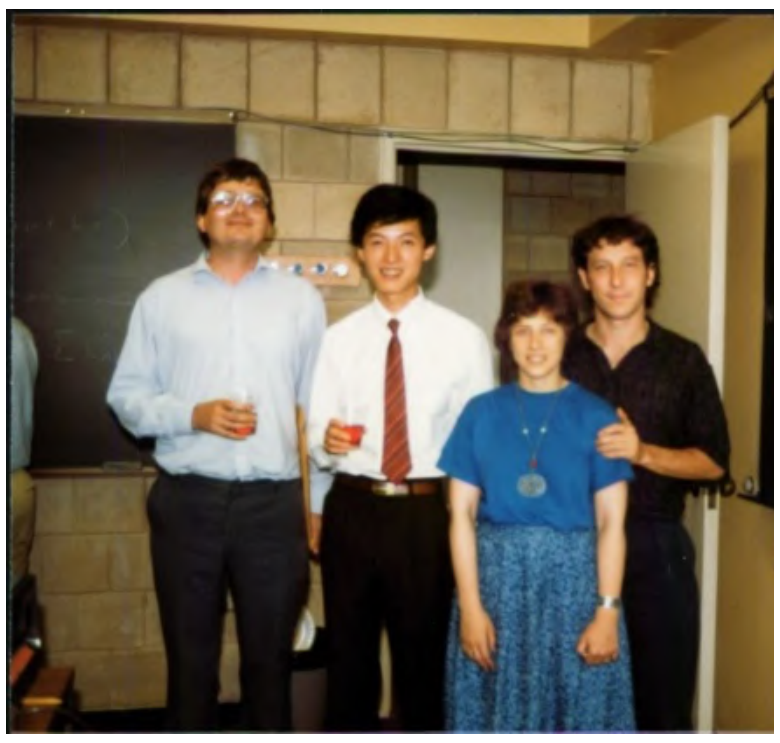

**Figure 11 Shoucheng with Per Bak, Steven Kivelson, and Steven's wife Pamela Davis at Stony Brook in 1987, celebrating Shoucheng's thesis defense.**

Along the same direction, in 1992 Shoucheng collaborated with Steven Kivelson and Dung-Hai Lee to propose a global phase diagram of the FQHE, which derived a set of interrelations among various FQHE states [6]. The Chern-Simons effective theory and the global phase diagram also revealed the relation between fractional quantum Hall physics and various forms of **particle-vortex duality**, which still remains an active research topic after 30 years.

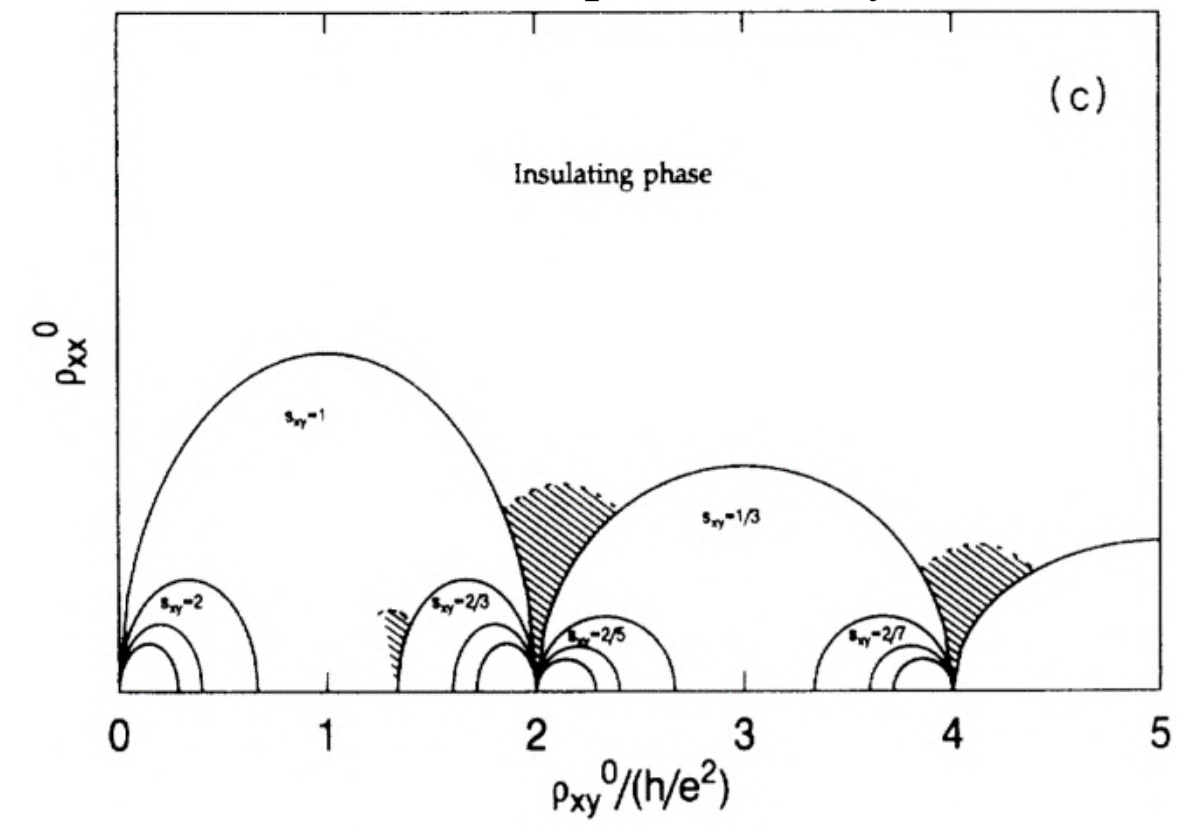

**Figure 12 Global phase diagram of FQHE, from Ref. [6].**

## 3. SO(5) Theory of High Temperature Superconductivity

### 3.1. *High $T_c$ cuprates and the SO(5) theory*

The discovery of high temperature superconductivity in copper oxides (known as cuprates) in 1987 was one of the most important breakthroughs in condensed matter physics. Despite enormous theoretical and experimental efforts, many questions about cuprates remain open today.

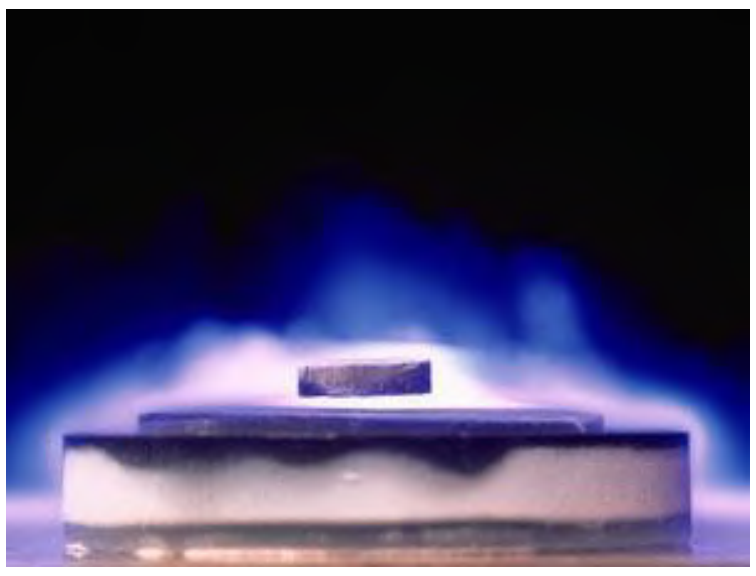

**Figure 13 A levitated magnet above superconducting $YBa_2Cu_3O_7$, from Science photo library**

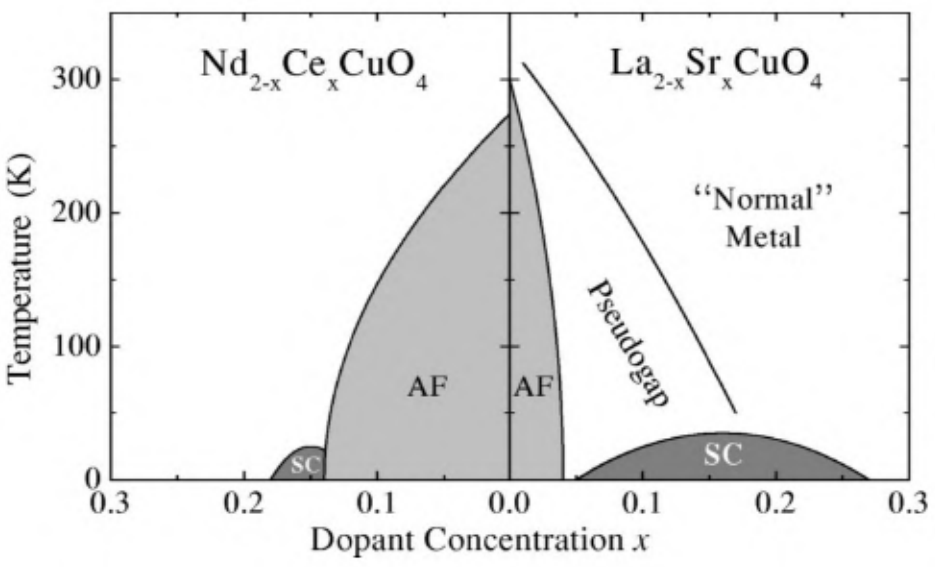

**Figure 14 Schematic phase diagram of Cuprates, from Ref. [19].**

Shoucheng started to work in this field in 1990, when he collaborated with C. N. Yang to point out that the Hubbard model, a model that is widely believed to capture the essential physics of cuprate superconductors, has an enhanced SO(4) symmetry [45]. Starting from this work, Shoucheng and collaborators explored the possibility of a (broken) enhanced symmetry in high $T_c$ superconductors and its physical consequences. For example, in 1995, Eugene Demler and Shoucheng studied the relation of a neutron scattering feature in cuprates with enhanced symmetry [46].

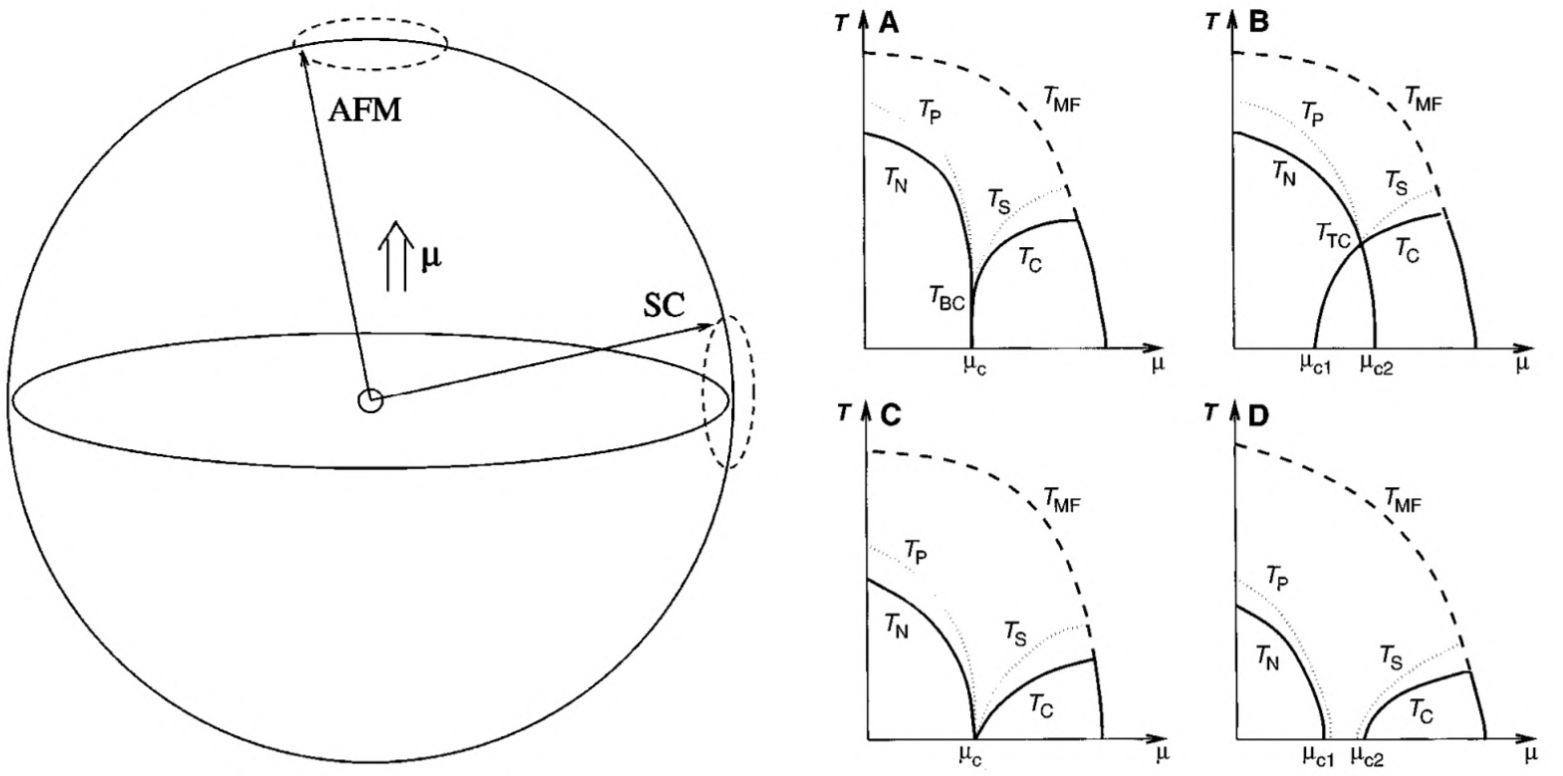

**Figure 15 (Left) The SO(5) superspin, from Ref. [4]. (Right) The SO(5) phase diagram, from Ref. [35].**

Based on these works, in 1997 Shoucheng proposed a new effective field theory for high $T_c$ superconductivity [35]. As an elegant application of symmetry principles, he proposed a five-component vector order parameter that incorporates both antiferromagnetism and superconductivity, the main two competing orders in the cuprates. This theory is known as the SO(5) theory, in which spin rotation symmetry SO(3) and charge conservation symmetry U(1)≃O(2) are considered as subgroups of SO(5). This theory offered a simple possible framework for interpreting many of the rich phenomena in cuprate superconductors, and it clearly reflected Shoucheng's unique style: relating simple and universal principles to experimental reality in condensed matter physics.

### 3.2. *Further developments*

Although the SO(5) theory was proposed as a phenomenological effective field theory, Shoucheng and collaborators also identified and studied microscopic models with an exact SO(5) symmetry [47, 48]. The latter paper further generalized the model to even larger symmetry, which is realizable in large-spin ultra-cold fermion systems [49].

A single electron carries the spinor representation of SO(5), in a manner analogous to the case of the spatial rotation group. Interestingly, the four-component SO(5) spinors are intrinsically related to the Dirac equation. Surprisingly, starting from these works, the SO(5) Clifford algebra and SO(5) spinors formed a theme that appeared again and again in Shoucheng's significant contributions in seemingly unrelated topics, including the four-dimensional quantum Hall effect, the intrinsic spin Hall effect, the quantum spin Hall effect, and topological insulators.

## 4. The Four-Dimensional Quantum Hall Effect

### 4.1. *From quaternions to quantum Hall effect in 4D*

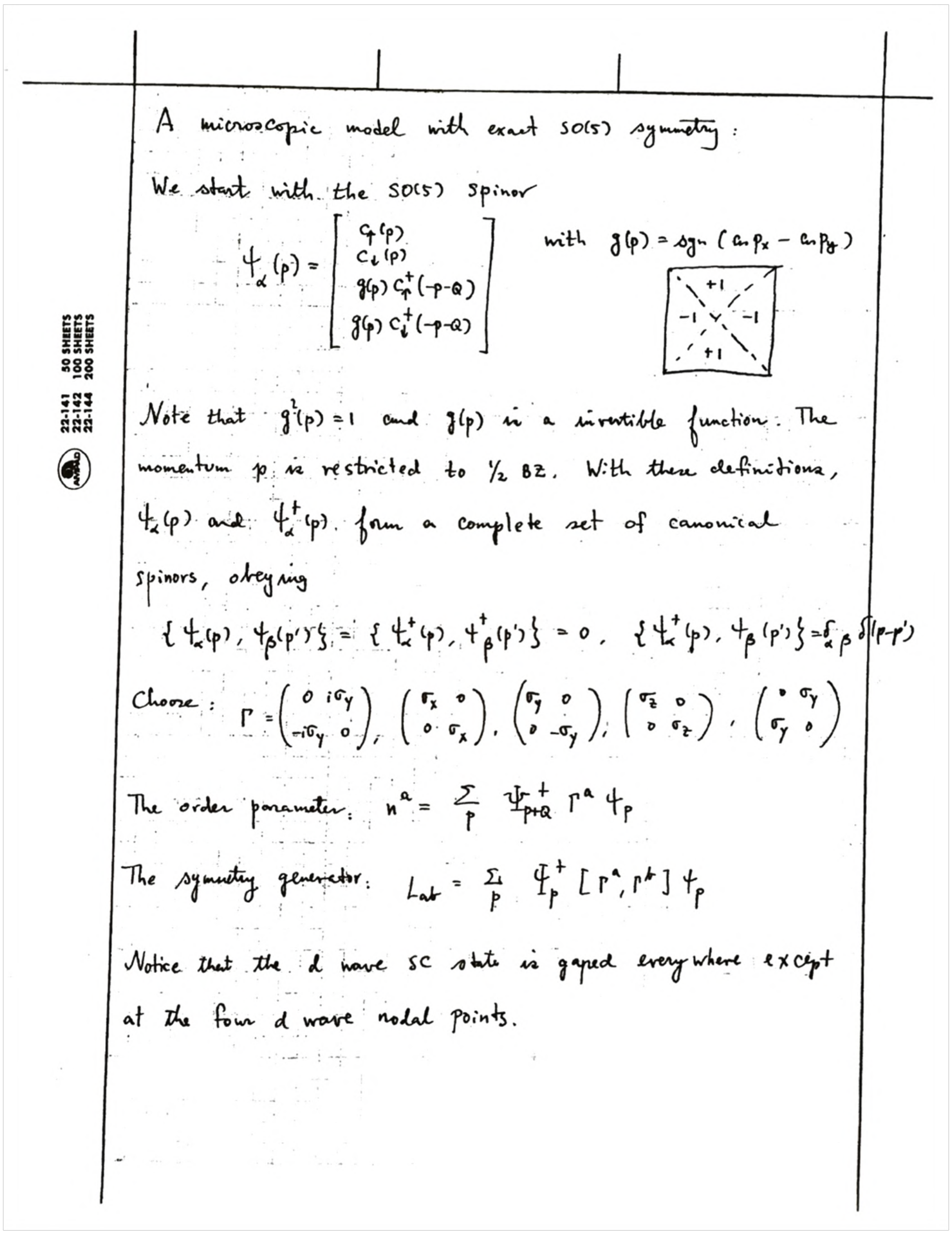

A microscopic model with exact SO(5) symmetry:

We start with the SO(5) spinor

$$\Psi_\alpha(p) = \begin{bmatrix} c_\uparrow(p) \\ c_\downarrow(p) \\ g(p)\, c^\dagger_\uparrow(-p-Q) \\ g(p)\, c^\dagger_\downarrow(-p-Q) \end{bmatrix} \quad \text{with } g(p) = \mathrm{sgn}(\cos p_x - \cos p_y)$$

Note that $g^2(p) = 1$ and $g(p)$ is a invertible function. The momentum $p$ is restricted to ½ BZ. With these definitions, $\Psi_\alpha(p)$ and $\Psi^\dagger_\alpha(p)$ form a complete set of canonical spinors, obeying

$$\{\Psi_\alpha(p), \Psi_\beta(p')\} = \{\Psi^\dagger_\alpha(p), \Psi^\dagger_\beta(p')\} = 0, \quad \{\Psi^\dagger_\alpha(p), \Psi_\beta(p')\} = \delta_{\alpha\beta}\,\delta(p-p')$$

Choose: $\Gamma = \begin{pmatrix} 0 & i\sigma_y \\ -i\sigma_y & 0 \end{pmatrix}, \begin{pmatrix} \sigma_x & 0 \\ 0 & \sigma_x \end{pmatrix}, \begin{pmatrix} \sigma_y & 0 \\ 0 & -\sigma_y \end{pmatrix}, \begin{pmatrix} \sigma_z & 0 \\ 0 & \sigma_z \end{pmatrix}, \begin{pmatrix} 0 & \sigma_y \\ \sigma_y & 0 \end{pmatrix}$

The order parameter: $n^a = \sum_p \Psi^\dagger_{p+Q} \Gamma^a \Psi_p$

The symmetry generator: $L_{ab} = \sum_p \Psi^\dagger_p [\Gamma^a, \Gamma^b] \Psi_p$

Notice that the d wave SC state is gaped everywhere except at the four d wave nodal points.

**Figure 16 Shoucheng's note about an SO(5) model in 1997, with Dirac Γ matrices.**

Shoucheng was deeply committed to the pursuit of mathematical beauty in physical theories, and his theory of four dimensional (4D) quantum Hall effect (QHE) [50] is one of the most vivid examples of the power of this approach.

A remarkable feature of the two-dimensional (2D) QHE is that the QHE wave functions can be written as functions of complex numbers $z = x + iy$, where $(x, y)$ represents the 2D coordinate of an electron. This motivated Shoucheng to ask: is there a QHE wave function of quaternions instead of complex numbers?

In mathematics, in addition to the real numbers $R$, and complex numbers $C$, there are also quaternions $H$ invented by William Rowan Hamilton, which have the form:

$$u = a + b\boldsymbol{i} + c\boldsymbol{j} + d\boldsymbol{k}\,,$$

with $\boldsymbol{i}^2 = \boldsymbol{j}^2 = \boldsymbol{k}^2 = -1$. Since a quaternion consists of 4 components, Shoucheng was led to consider the 4D QHE with 4 coordinates. In theoretical physics, $\boldsymbol{i}, \boldsymbol{j}, \boldsymbol{k}$ can be written as the $2 \times 2$ Pauli matrices $i\sigma_1$, $i\sigma_2, i\sigma_3$, which are the generators of the spin-1/2 representation of the SU(2) group. Therefore, the 4D QHE quaternion wave function couples to an SU(2) Yang-Mills gauge field, instead of the U(1) gauge field in 2D quantum Hall states.

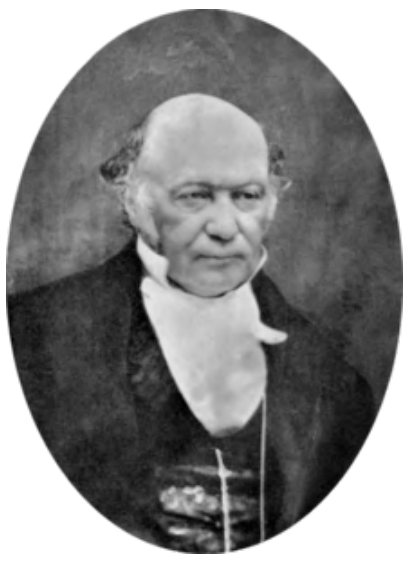

**Figure 17 William Rowan Hamilton invented quaternions**

The original paper of Shoucheng and his student Jiangping Hu on the 4D QHE [50] (2001) is a beautiful synthesis of mathematics and physics. They employed the 2nd Hopf map, a topological map from the 7D sphere to the 4D sphere, to construct the 4D QHE quaternion wave function. This revealed that the 4D QH state is a topological state with a nonzero 2nd Chern number, which leads to an SU(2) quantized Hall effect.

## 4.2. *Relation to more recent developments*

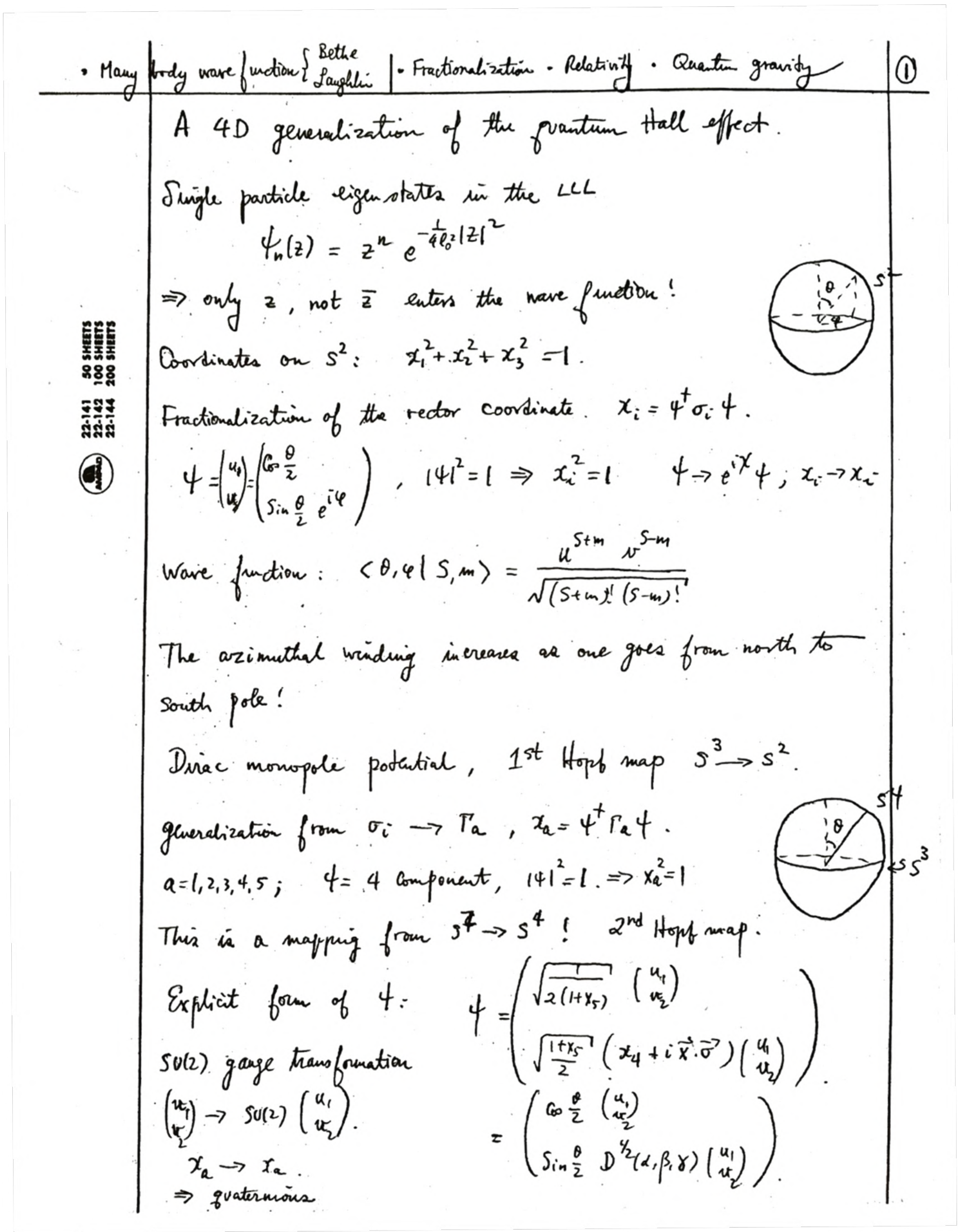

• Many body wave function { Bethe / Laughlin | • Fractionalization • Relativity • Quantum gravity ①

A 4D generalization of the quantum Hall effect.

Single particle eigenstates in the LLL

$\psi_n(z) = z^n e^{-\frac{1}{4\ell_0^2}|z|^2}$

⇒ only $z$, not $\bar{z}$ enters the wave function!

Coordinates on $S^2$: $x_1^2 + x_2^2 + x_3^2 = 1$.

Fractionalization of the vector coordinate. $x_i = \psi^\dagger \sigma_i \psi$.

$\psi = \begin{pmatrix} u_1 \\ u_2 \end{pmatrix} = \begin{pmatrix} \cos\frac{\theta}{2} \\ \sin\frac{\theta}{2} e^{i\varphi} \end{pmatrix}$, $|\psi|^2 = 1 \Rightarrow x_i^2 = 1$ $\psi \to e^{i\chi}\psi$; $x_i \to x_i$

Wave function: $\langle \theta, \varphi | S, m \rangle = \frac{u^{S+m} v^{S-m}}{\sqrt{(S+m)!(S-m)!}}$

The azimuthal winding increases as one goes from north to south pole!

Dirac monopole potential, 1st Hopf map $S^3 \to S^2$.

generalization from $\sigma_i \to \Gamma_a$, $x_a = \psi^\dagger \Gamma_a \psi$.

$a = 1,2,3,4,5$; $\psi$ = 4 component, $|\psi|^2 = 1 \Rightarrow x_a^2 = 1$

This is a mapping from $S^7 \to S^4$! 2nd Hopf map.

Explicit form of $\psi$: $\psi = \begin{pmatrix} \sqrt{\frac{1}{2(1+x_5)}} \begin{pmatrix} u_1 \\ u_2 \end{pmatrix} \\ \sqrt{\frac{1+x_5}{2}} (x_4 + i\vec{x}\cdot\vec{\sigma}) \begin{pmatrix} u_1 \\ u_2 \end{pmatrix} \end{pmatrix} = \begin{pmatrix} \cos\frac{\theta}{2} \begin{pmatrix} u_1 \\ u_2 \end{pmatrix} \\ \sin\frac{\theta}{2} D^{1/2}(\alpha,\beta,\gamma) \begin{pmatrix} u_1 \\ u_2 \end{pmatrix} \end{pmatrix}$.

SU(2) gauge transformation

$\begin{pmatrix} u_1 \\ u_2 \end{pmatrix} \to SU(2) \begin{pmatrix} u_1 \\ u_2 \end{pmatrix}$.

$x_a \to x_a$.

⇒ quaternions

**Figure 18 Shoucheng's note in 2001 on 4D QHE**

Noting that the 4D QH state has edge states consisting of relativistic particles with different spins living in 3D space, Zhang and Hu proposed it as a candidate

"unified theory" of fundamental interactions and particles. This theory is also one of the earliest studies of topological states of matter in high dimensions, and its idea has been generalized in many directions, including in numerous studies of other high dimensional topological states.

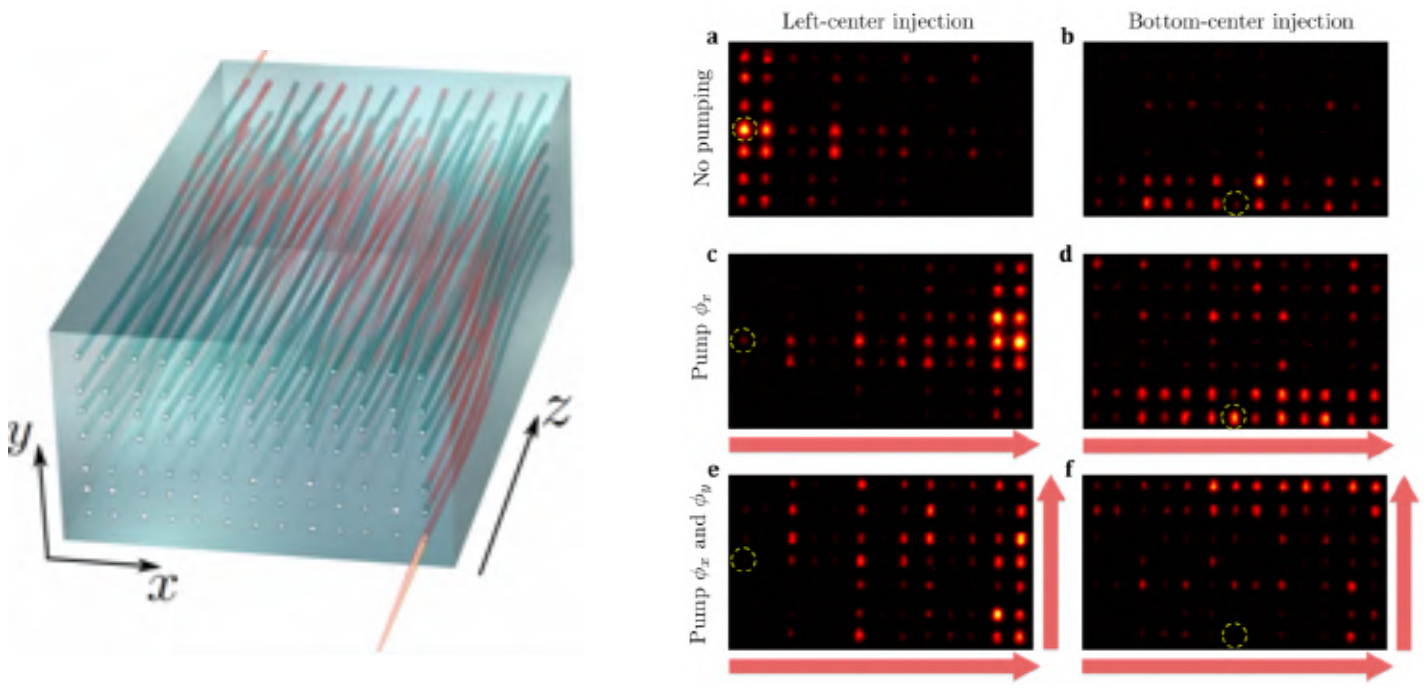

**Figure 19 4D QHE edge state pumping observed in the photonic crystal experiment [14].**

The 4D QHE theory partially motivated Shoucheng's later work on the quantum spin Hall (QSH) effect and topological insulators (TI). Zhang, Hu (2001) [50] proposed the following equation of a spin Hall current, as a consequence of the 4D QHE:

$$\dot{X}_a = \frac{R^4 \partial V}{I^2 \partial X_b} F^i_{ab} I_i, \dot{I}_i = \epsilon_{ijk} A^j_\mu \dot{X}_\mu I_k$$

The spin Hall current of electrons corresponds to the SU(2) Hall current of the 4D. In 2008, work from his group clarified that the 3D TI and 2D QSH can be obtained from a 4D QH state (though a different one from the original Zhang-Hu proposal) via dimensional reduction [3].

## 4.3. *Experimental realization*

Since we only have 3 spatial dimensions in reality, one might not expect a 4D QHE to be experimentally accessible. However, in 2018 experiments on photonic crystal [14] and cold atom systems [51] effectively "realized" the 4D QHE. The key idea is to use control parameters to create synthetic dimensions, much in the way Thouless pumping in 1D is related to the 2D QHE. These experiments effectively constructed a 4D system with a non-zero $2^{nd}$ Chen number by introducing two pumping parameters tunable by laser beams, together with two spatial dimensions.

# 5. Spin Hall Effect

## 5.1. *Intrinsic spin Hall effect*

Shoucheng's work on the 4D QH effect implied the possibility of topological and dissipationless spin transport in 3D systems. Although the original 4D QH model is not directly related to any experimental system, this connection motivated Shoucheng to start thinking about spin-orbit coupling and its role in dissipationless transport. Near the end of the Zhang-Hu paper, the authors commented that “The single-particle states also have a strong gauge coupling between iso-spin and orbital degrees of freedom, which is ultimately responsible for the emergence of the *relativistic helicity* of the collective modes.” In 2003, Shuichi Murakami, Naoto Nagaosa and Shoucheng predicted an intrinsic spin Hall effect in semiconductors [8], which referred to a transverse dissipationless spin current induced by an electric field in semiconductor compounds with strong spin-orbit coupling. The spin Hall current is described by the following formula:

$$j_j^i = \sigma_s \epsilon^{ijk} E_k$$

As was stated at the beginning of their paper, this work “is driven by the confluence of the important technological goals of quantum spintronics with the quest of generalizing the quantum Hall effect (QHE) to higher dimensions.”

Different from charge current, spin current is even under time reversal, which

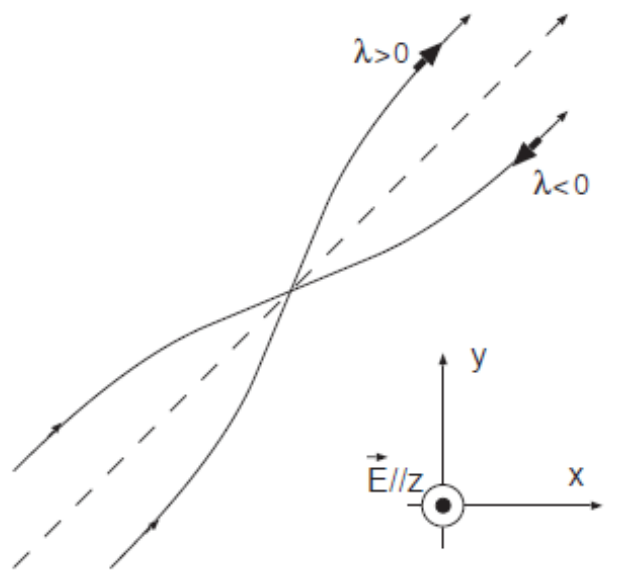

**Figure 20 Schematic illustration of the mechanism of intrinsic spin Hall effect in Ref. [8].**

allows this effect to occur in materials without a magnetic field. Spin-orbit coupling replaced the role of magnetic field in QHE and leads to the nontrivial Berry's phase that is the essential reason of the spin Hall effect. The 4DQH and intrinsic spin Hall effect are the first proposals of dissipationless transport in dimensions higher than two, which is the overture of the upcoming breakthrough of new topological insulators and topological superconductors. (Intrinsic spin Hall effect of a different type was also proposed in the concurrent work of [52])

Soon after its theoretical proposal, the spin Hall effect was observed experimentally in hole doped semi-conductors [28].

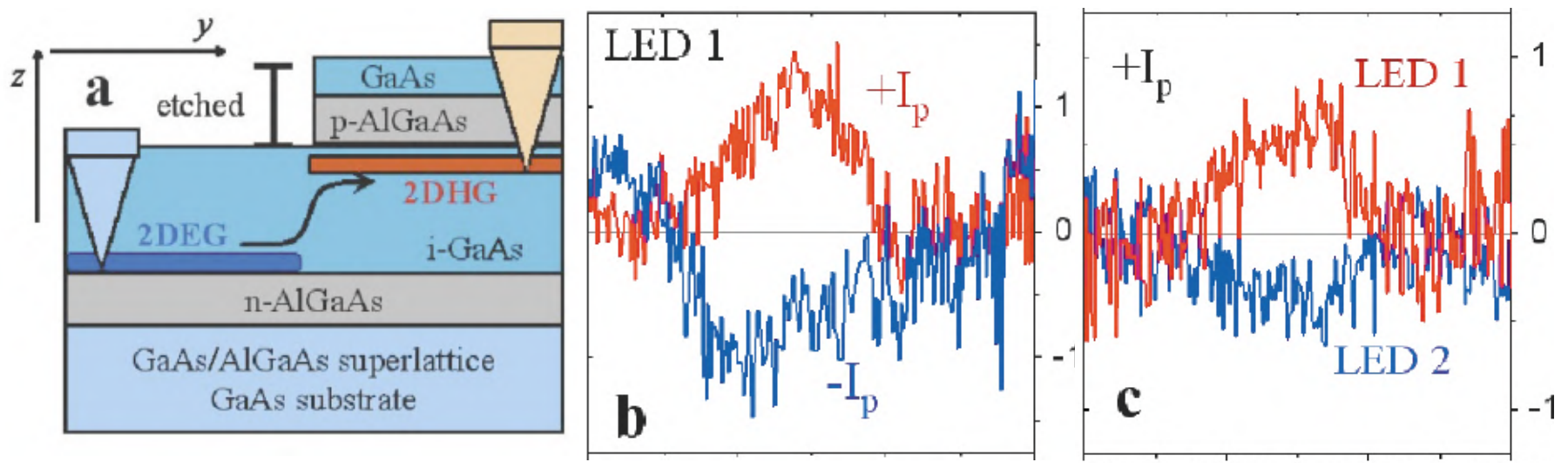

**Figure 21 (a) Experimental device and (b) observed edge spin accumulation (by measurement of light polarization) due to spin Hall effect in a 2D hole gas system from Ref. [28].**

## 5.2. *Spin Hall insulators*

In 2004, Murakami, Nagaosa, and Shoucheng generalized their results to a family of materials that they named "spin Hall insulators," which are systems with zero charge conductivity but finite spin Hall conductivity [7]. The materials mentioned in this paper are zero gap and narrow gap semiconductors, including HgTe, HgSe, $\beta$-HgS, $\alpha$-Sn, PbTe, PbSe, PbS. Many of the materials mentioned here were later identified as topological insulators, although the connection to topological physics had not yet been uncovered in this work.

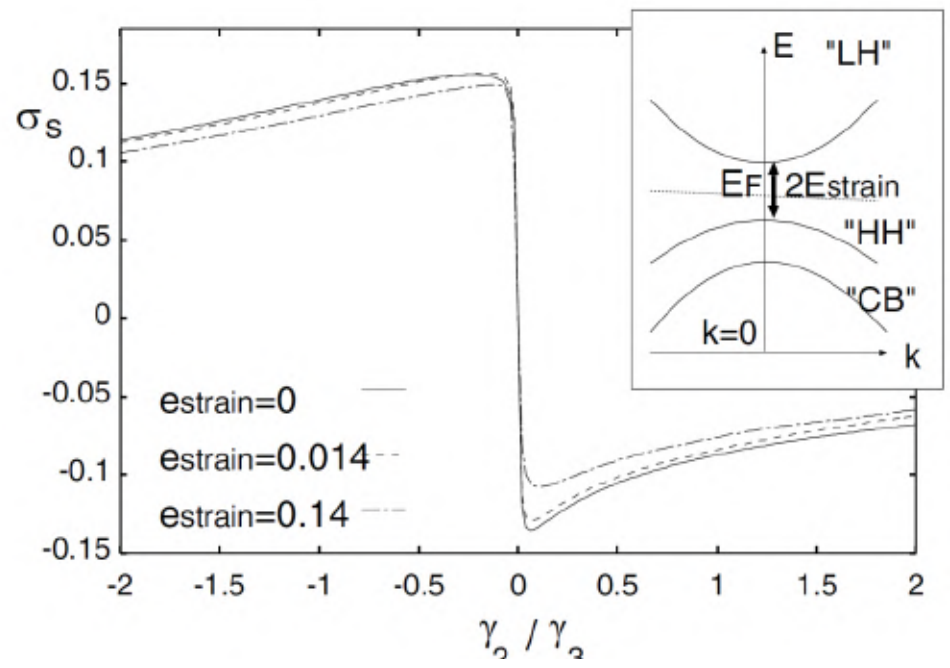

**Figure 22 Spin Hall conductivity and band structure of the spin Hall insulator [7].**

### 5.3. *Other works in spintronics*

In addition to the spin Hall effect, Shoucheng also worked more broadly in the field of spintronics. For example, in his collaboration with B. A. Bernevig and J. Orenstein [53], they predicted, and then subsequently experimentally realized, a "persistent spin helix" which supported a long-lived helical spin configuration due to an emergent SU(2) symmetry. Shoucheng's works brought many fundamental physics ideas into the field of semiconductors and spintronics.

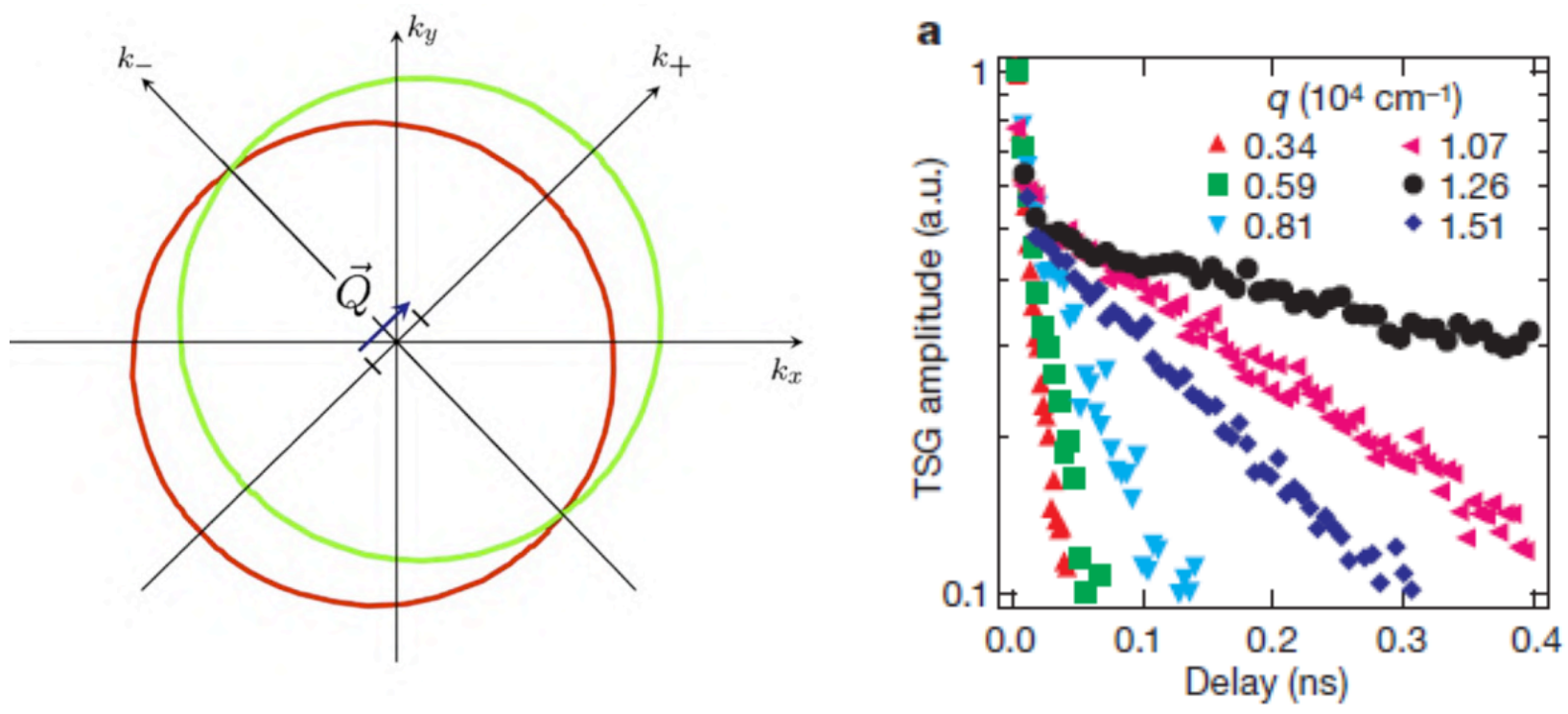

**Figure 23 (Left) Theoretical prediction of the special Fermi surface that leads to persistent spin helix. (Right) Experimental realization of the spin helix in Ref. [25].**

## 6. Quantum spin Hall effect

### 6.1. *Early models of quantum spin Hall effect*

The 4D QH and the intrinsic spin Hall effect are both proposed as generalizations of the quantum Hall effect. In 2004, Shoucheng started to consider a more direct generalization of the QHE. B. A. Bernevig and Shoucheng proposed that in a semiconductor with a spatially inhomogeneous strain, the spin-orbit coupling can realize a uniform magnetic field that is opposite for spin up and spin down electrons. This manifests as a quantized spin Hall (QSH) effect preserving time reversal symmetry [15].

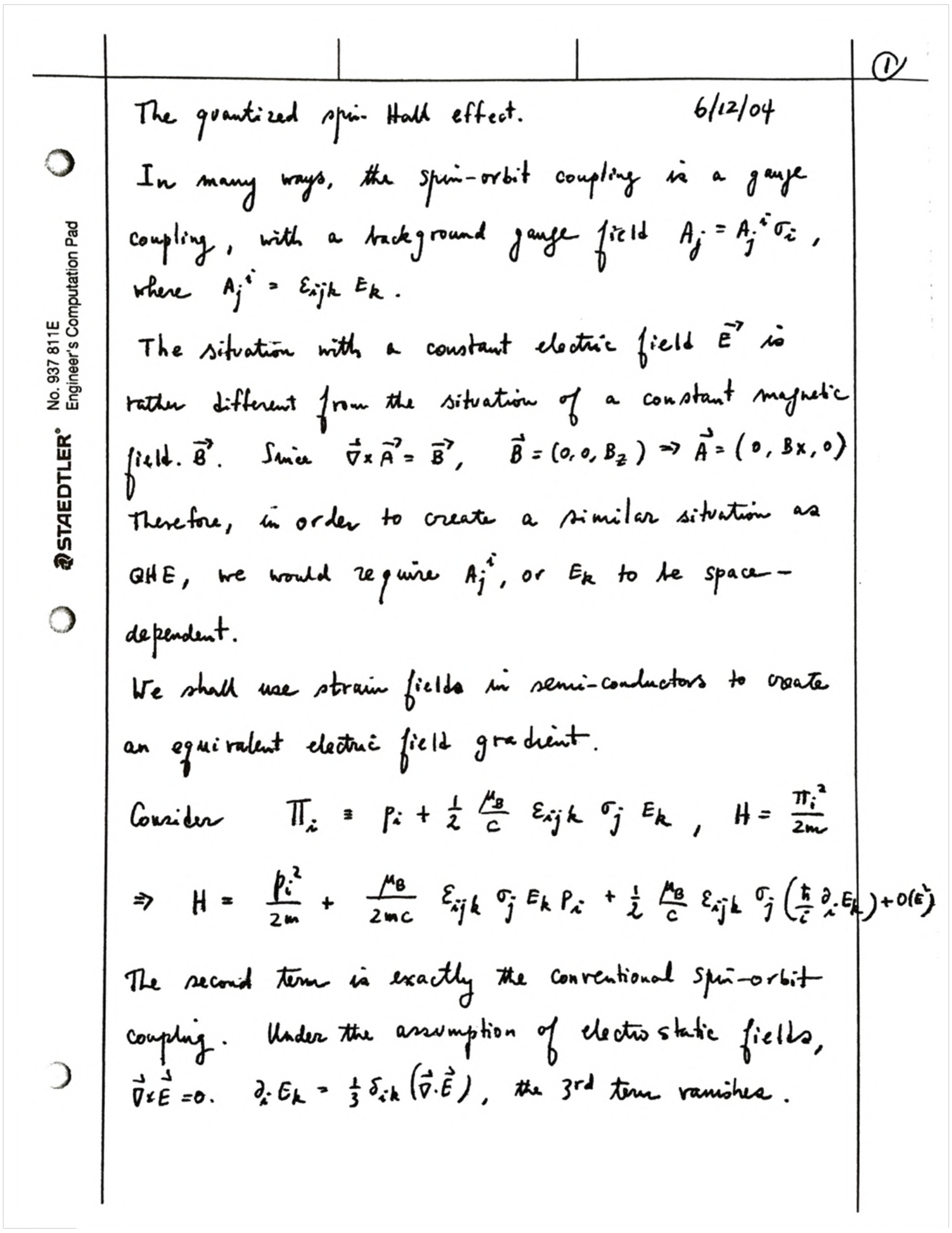

STAEDTLER® No. 937 811E Engineer's Computation Pad

①

The quantized spin Hall effect. 6/12/04

In many ways, the spin-orbit coupling is a gauge coupling, with a background gauge field $A_j = A_j^i \sigma_i$, where $A_j^i = \varepsilon_{ijk} E_k$.

The situation with a constant electric field $\vec{E}$ is rather different from the situation of a constant magnetic field. $\vec{B}$. Since $\vec{\nabla} \times \vec{A} = \vec{B}$, $\vec{B} = (0, 0, B_z) \Rightarrow \vec{A} = (0, Bx, 0)$

Therefore, in order to create a similar situation as QHE, we would require $A_j^i$, or $E_k$ to be space-dependent.

We shall use strain fields in semi-conductors to create an equivalent electric field gradient.

Consider $\Pi_i \equiv p_i + \frac{1}{2} \frac{\mu_B}{c} \varepsilon_{ijk} \sigma_j E_k$, $H = \frac{\Pi_i^2}{2m}$

$$\Rightarrow H = \frac{p_i^2}{2m} + \frac{\mu_B}{2mc} \varepsilon_{ijk} \sigma_j E_k p_i + \frac{1}{2} \frac{\mu_B}{c} \varepsilon_{ijk} \sigma_j \left(\frac{\hbar}{i} \partial_i E_k\right) + O(E^2)$$

The second term is exactly the conventional spin-orbit coupling. Under the assumption of electrostatic fields, $\vec{\nabla} \times \vec{E} = 0$. $\partial_i E_k = \frac{1}{3} \delta_{ik} (\vec{\nabla} \cdot \vec{E})$, the 3rd term vanishes.

**Figure 24 Shoucheng's note in 2004 on quantized spin Hall effect.**

Concurrently, a different quantum spin Hall model was independently proposed by Charles L. Kane and J. Eugene Mele, who studied spin-orbit coupling in graphene [31]. This model is a time-reversal invariant generalization of a model proposed by F. D. M. Haldane in 1988 [54]. In another work soon afterwards, Kane and Mele pointed out that the quantum spin Hall state is classified by $Z_2$, which means that the states with odd pairs of edge states are topologically robust in the presence of time-reversal symmetry [55].

Shoucheng and his students Congjun Wu and B. Andrei Bernevig studied the stability of QSH edge states in the presence of electron interaction. They pointed out that the interacting edge states are qualitatively different from the regular Luttinger liquid, and named it the "helical liquid" [56] (see also [57]).

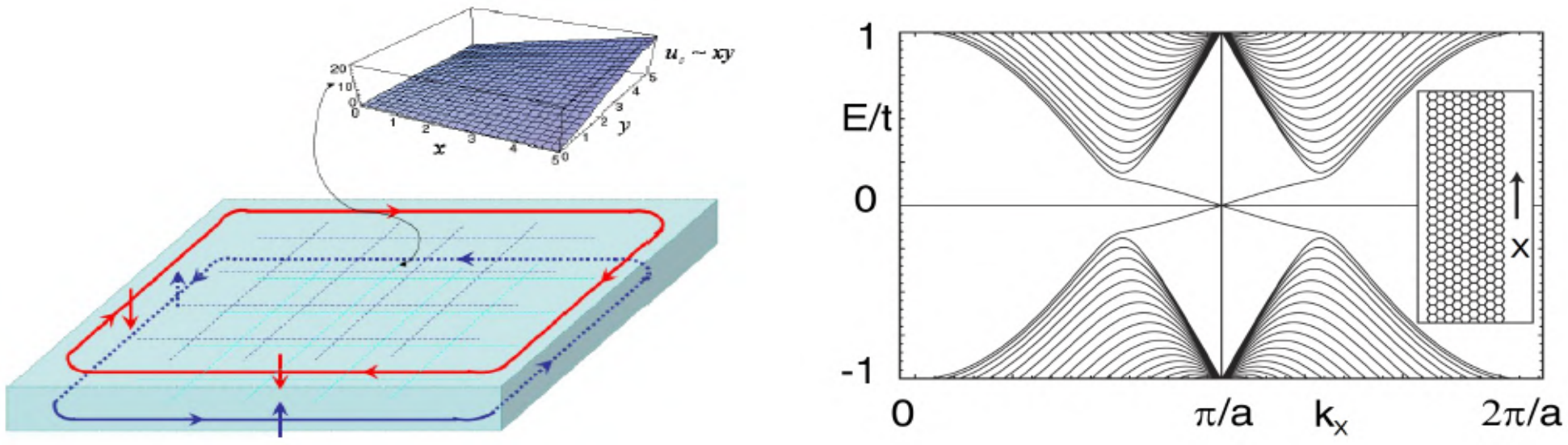

**Figure 25 The two earliest models of the quantum spin Hall effect. (Left) The illustration of the Bernevig-Zhang proposal [15]. (Right) The energy spectrum of the Kane-Mele model [31].**

### 6.2. *HgTe theory and experiments*

The earliest models reviewed above set up the theoretical foundation of QSH, but they are difficult to realize experimentally for various reasons. In 2006, Shoucheng and his students B. A. Bernevig and T. L. Hughes made the first realistic proposal of a QSH material: HgTe/CdTe quantum wells [5]. They obtained the low energy effective theory of this quantum well (known as the Bernevig-Hughes-Zhang (BHZ) model) and predicted that the QSH phase could be identified by a topological phase transition controlled by the thickness of the quantum well. The BHZ work not only proposed the first realistic QSH material, but also proposed a general mechanism to identify topological materials, which they named "band inversion."

Soon after the proposal, the quantum spin Hall effect in HgTe quantum wells was verified in Laurens Molenkamp's group [16]. They indeed observed a $2e^2/h$ conductance for $d>d_c$ while the resistance is much higher when $d<d_c$, agreeing well with theoretical predictions.

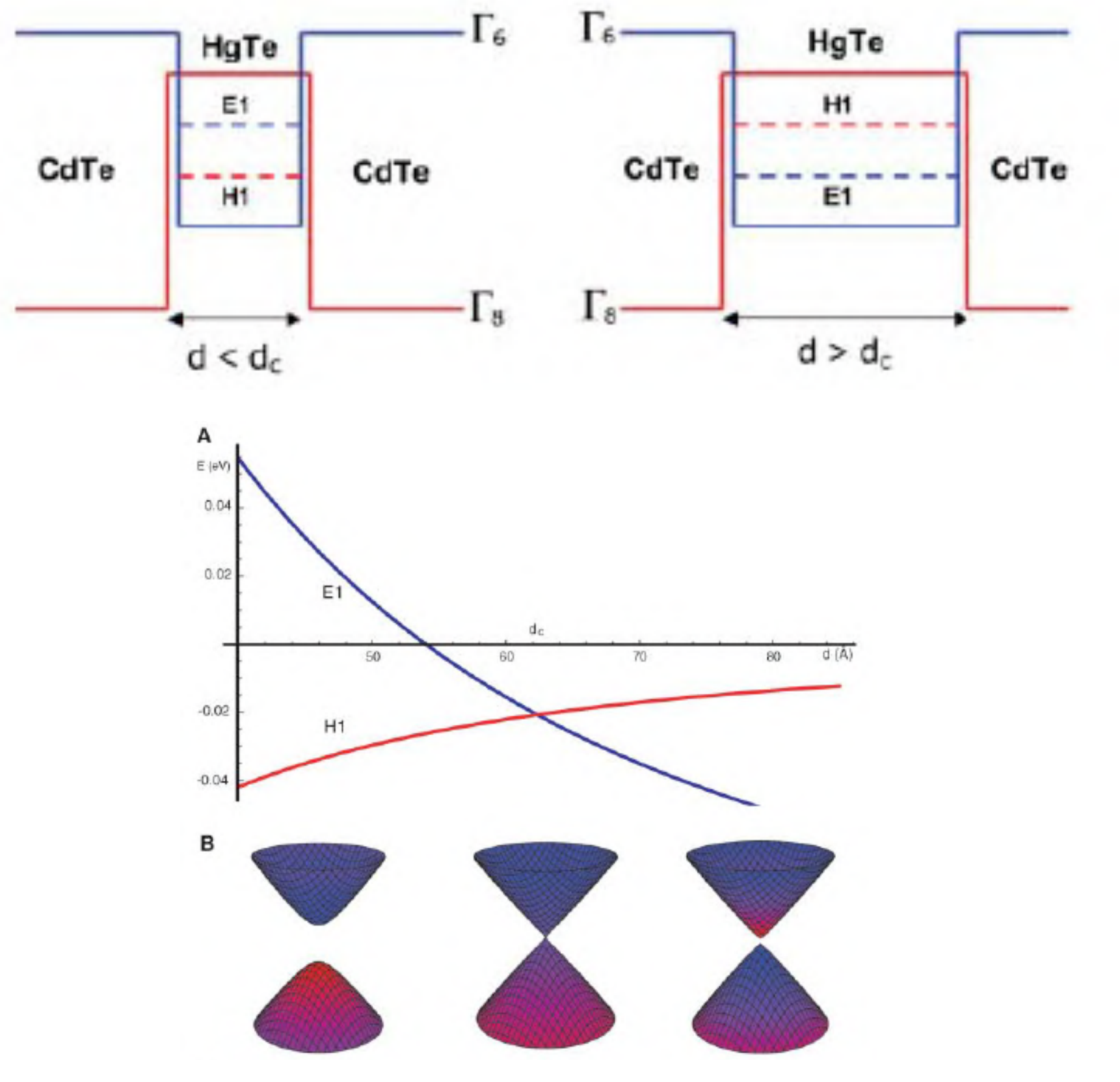

**Figure 26 The band edge energy of narrow and wide quantum wells [5].**

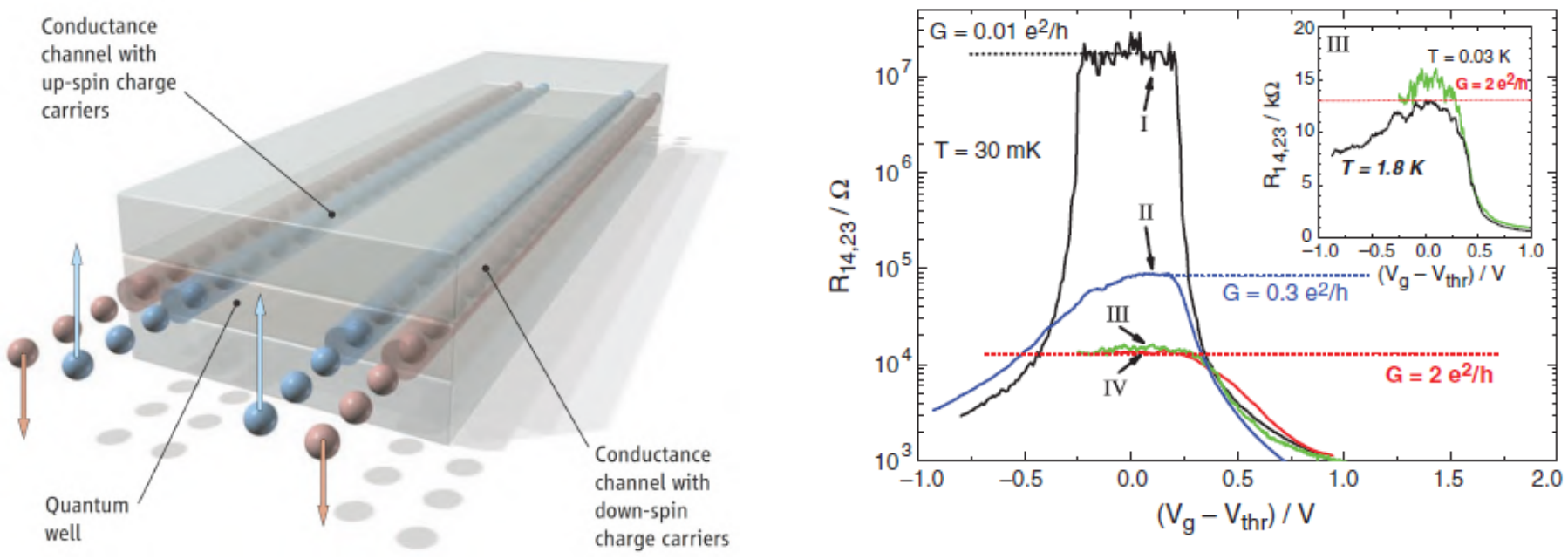

**Figure 27 (Left) Illustration of the quantum spin Hall effect. (Right) Transport measurement in HgTe quantum wells, from Ref. [16].**

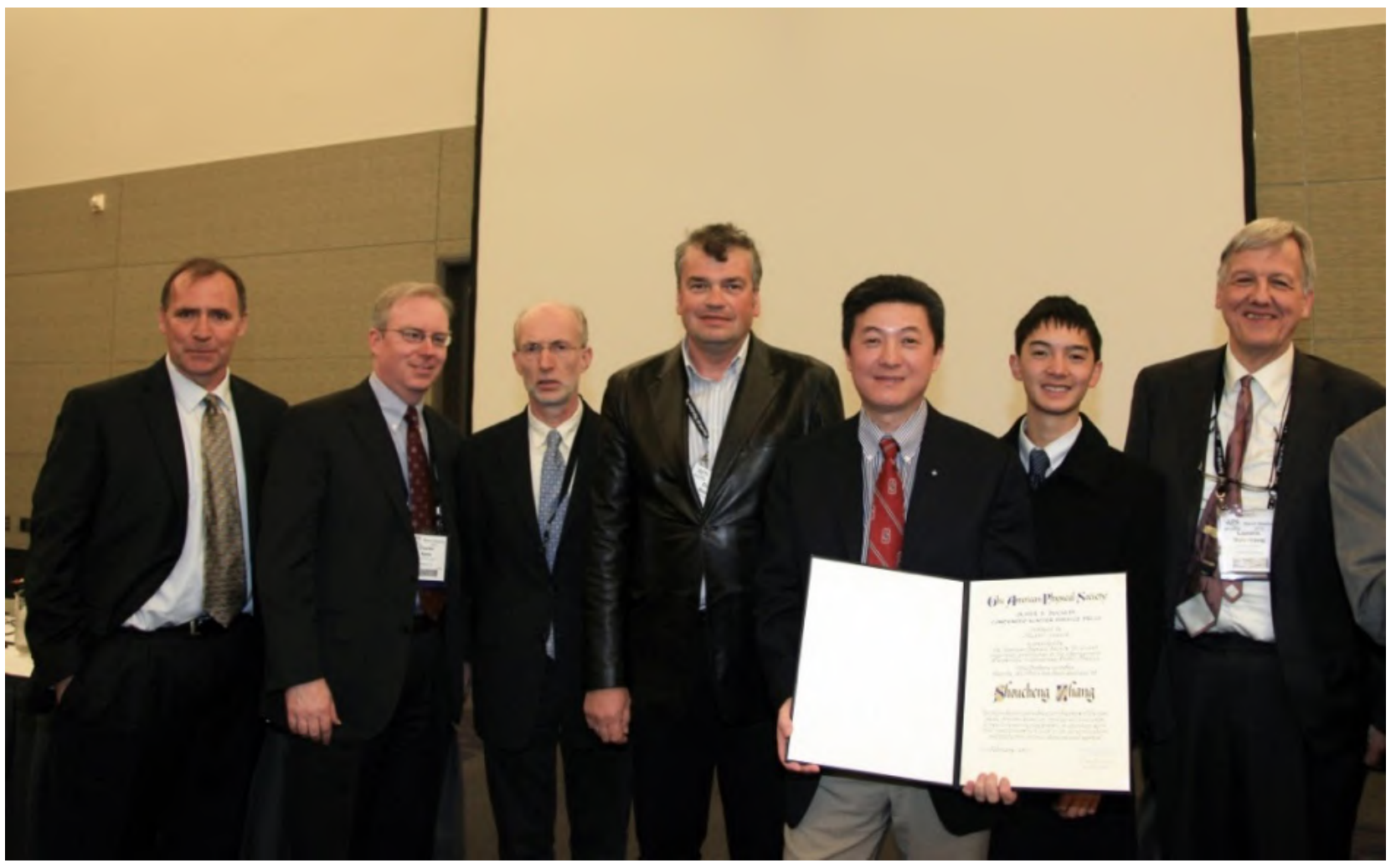

**Figure 28 Shoucheng (5th from left), Charles L. Kane (2nd) and Laurens Molenkamp (7th) at the Oliver E. Buckley Prize ceremony (2012), together with Shoucheng's son Brian (6th) and colleagues Eric Fullerton (1st), Stuart S. P. Parkin (3rd), and Dimitri Basov (4th**

## 6.3. *Other quantum spin Hall materials*

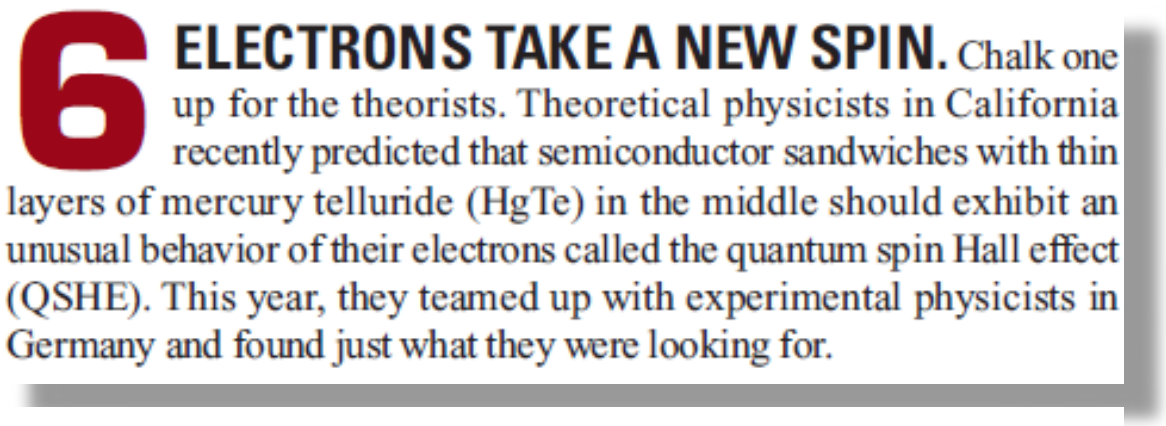
6 ELECTRONS TAKE A NEW SPIN. Chalk one up for the theorists. Theoretical physicists in California recently predicted that semiconductor sandwiches with thin layers of mercury telluride (HgTe) in the middle should exhibit an unusual behavior of their electrons called the quantum spin Hall effect (QSHE). This year, they teamed up with experimental physicists in Germany and found just what they were looking for.

**Figure 29 Dicovery of the quantum spin Hall effect was listed by Science magazine as one of the top 10 Breakthroughs of Year in 2007.**

The discovery of HgTe QSH heralded the beginning of a new era with tremendous theoretical and experimental developments in topological states of matter. Since then Shoucheng and collaborators have also predicted several other QSH materials, such as InAs/GaSb quantum wells [9] and stanene, a single-layer thin film of Sn [13].

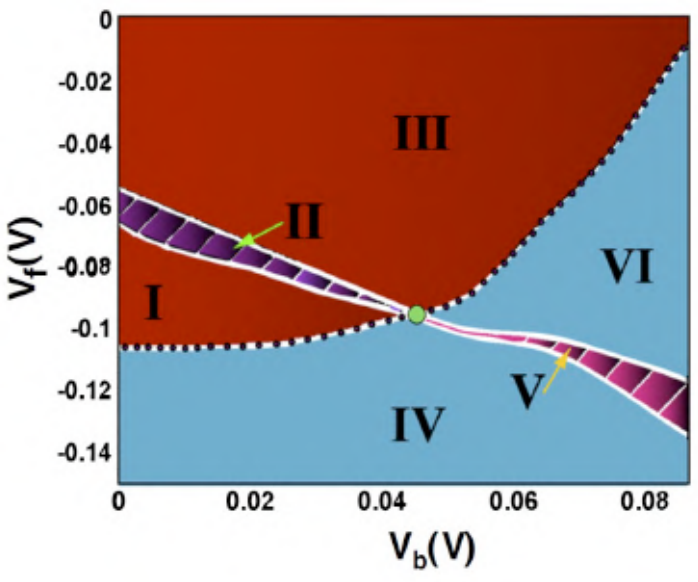

**Figure 30 Phase diagram of the InAs/GaSb quantum wells with QSH effect, from Ref. [9].**

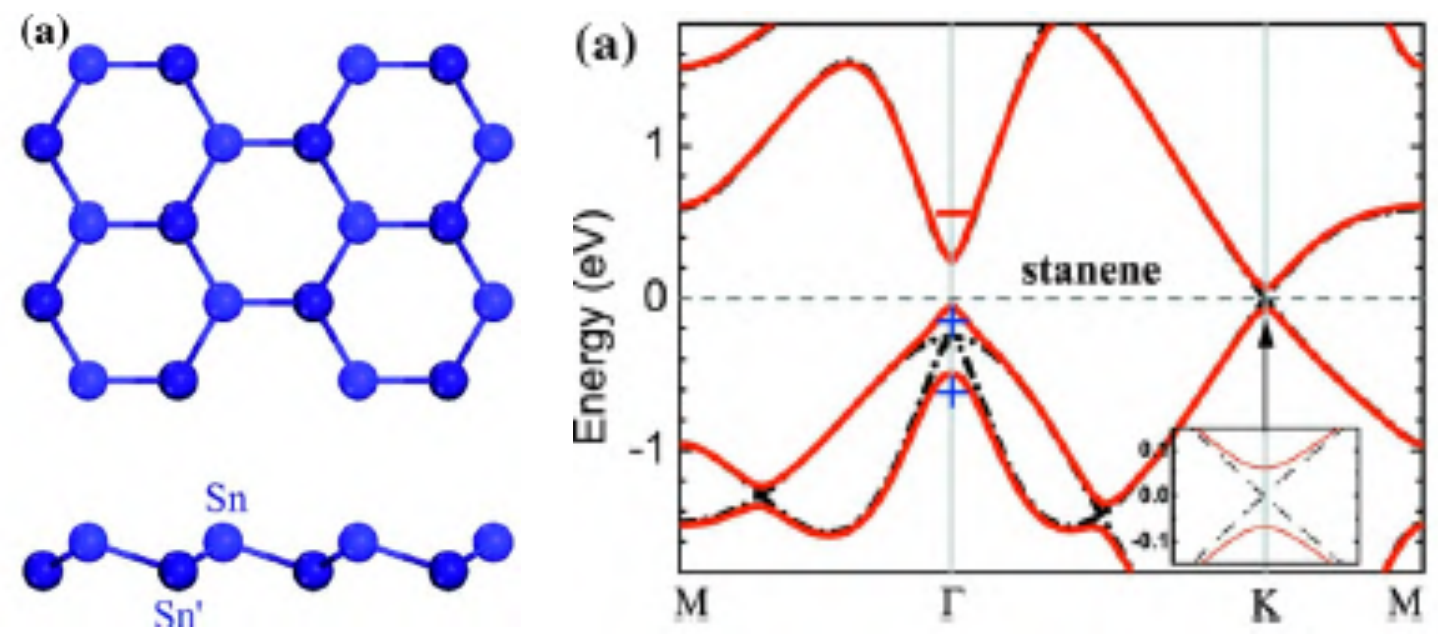

**Figure 31 Stanene and its non-trivial band structure, from Ref. [13].**

Shoucheng's group has also investigated many topological effects in QSH systems, such as electron interaction effects on the edge, and fractional charge on an edge magnetic domain wall, spin-charge separation induced by a $\pi$ flux, etc.

## 7. Three dimensional topological insulators

### 7.1. *Generalization of quantum spin Hall effect to three dimensions*

The understanding of the QSH and $Z_2$ topological protection implied that the QSH could be generalized to three dimensional insulators, with robust two-dimensional surface states. The surface electrons are "helical", with their spin direction locked with the momentum. This new state, named 3D topological insulators (TI), was proposed in 2006 by three papers [58-60].

The first 3D TI was the alloy of Bi and Sb, proposed by L. Fu and C. L. Kane [61], and realized by M. Z. Hasan's group in Princeton [62].

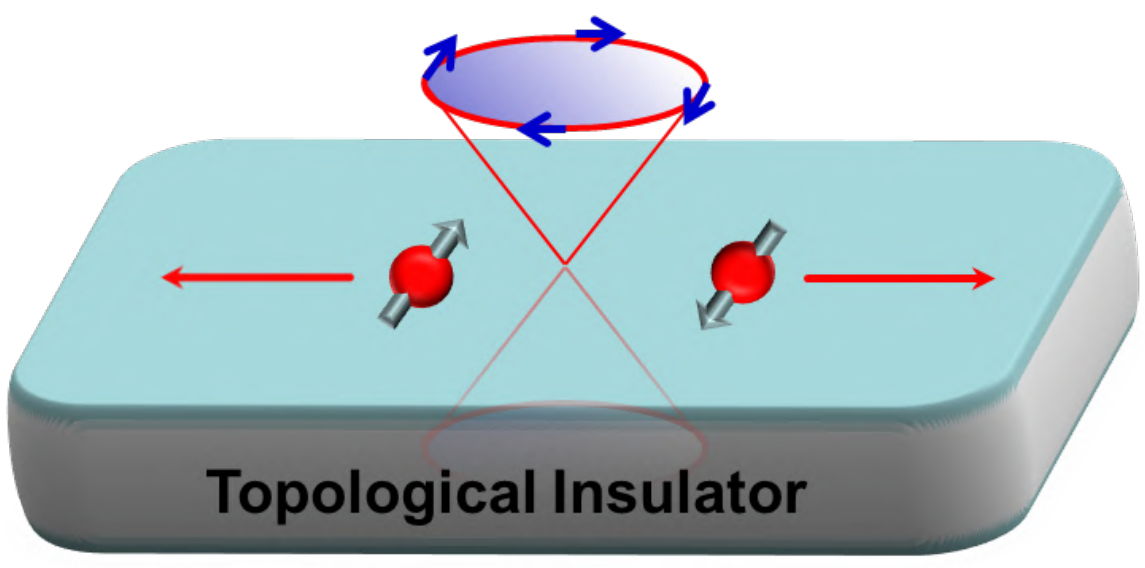

**Figure 32 An illustration of the spin-momentum locking of topological insulator surface states.**

In 2008, Shoucheng and collaborators proposed a new family of topological insulators, including three materials: $Bi_2Te_3$, $Bi_2Se_3$, and $Sb_2Te_3$ [63]. These materials are described by a simple effective model similar to the BHZ model. This family of TI was realized experimentally in 2009 [33]. The $Bi_2Se_3$ family of TI is easy to grow and has a large gap and simple surface states, which immediately became the "gold standard" of TI. They remain the most widely studied TI materials to this day.

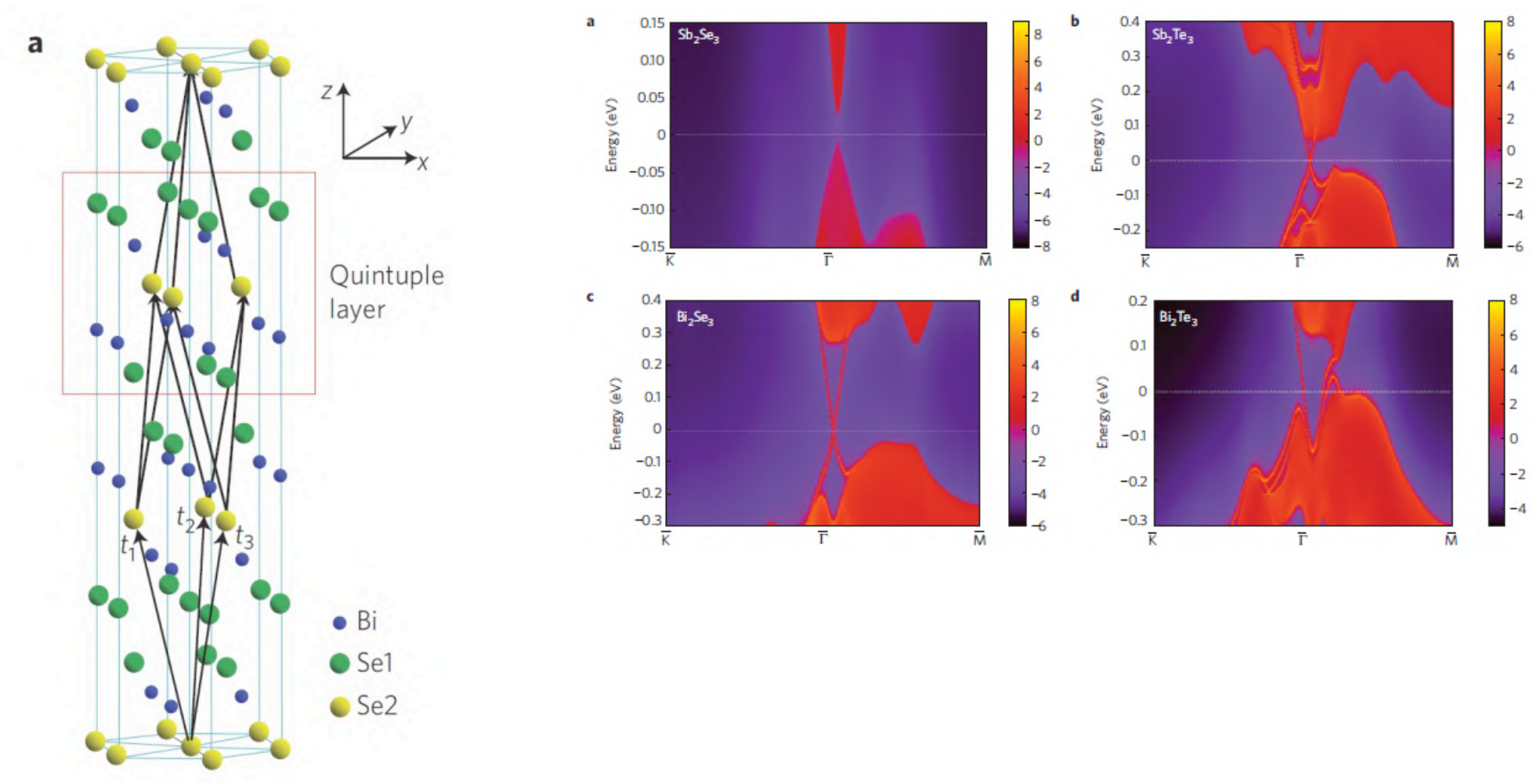

**Figure 33 (Left) Crystal structure of the Bi2Se3 family of TI. (Right) Theoretical prediction of the surface state dispersion relation based on ab initio calculation. From Ref. [17].**

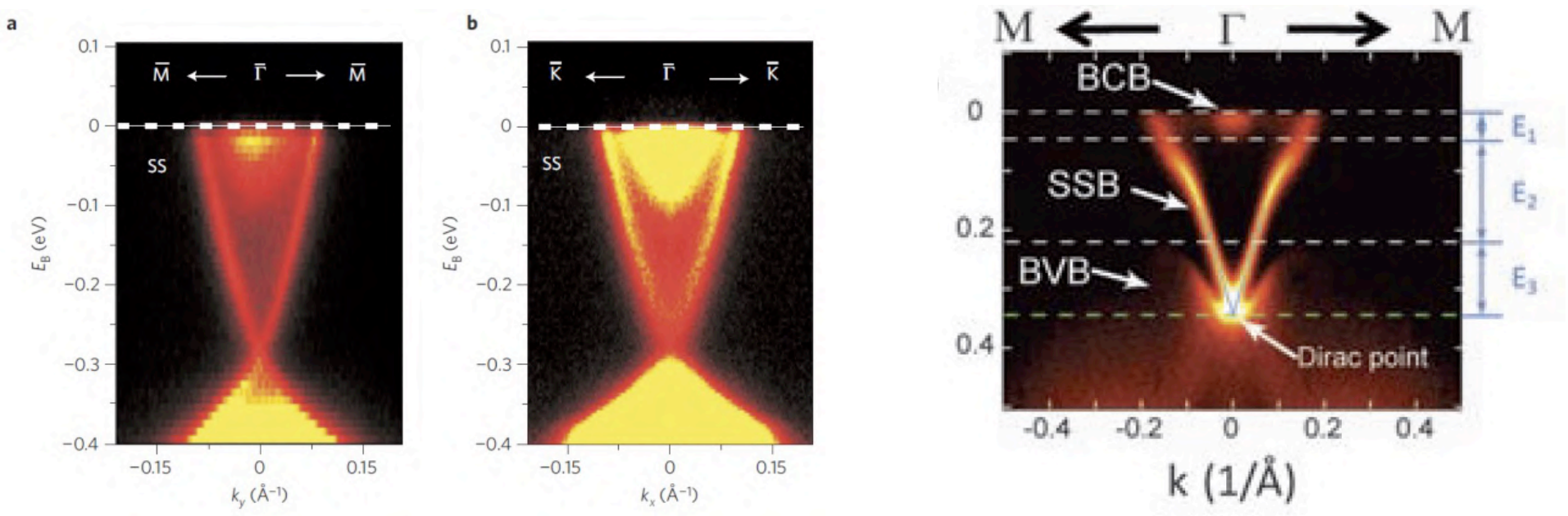

**Figure 34 Experimental observation of topological surface states in angle-resolved photoemission. (Left) Bi2Se3 from Ref. [29]. (Right) Bi2Te3 from Ref. [33].**

## 7.2. *Other 3D TI materials*

Since 3D TI's are single crystals rather than quantum wells, it is easier to realize experimentally. Shoucheng and collaborators have predicted many of the early 3D TI materials. This includes strained HgTe [64] (Xi Dai et al., *Phys. Rev. B* **77**, 125319 (2008), realized in L. Molenkamp's group), half-Heusler compounds [21], $TlBiSe_2$ family [65], filled skutterudites [66], Actinide compounds [67], etc.

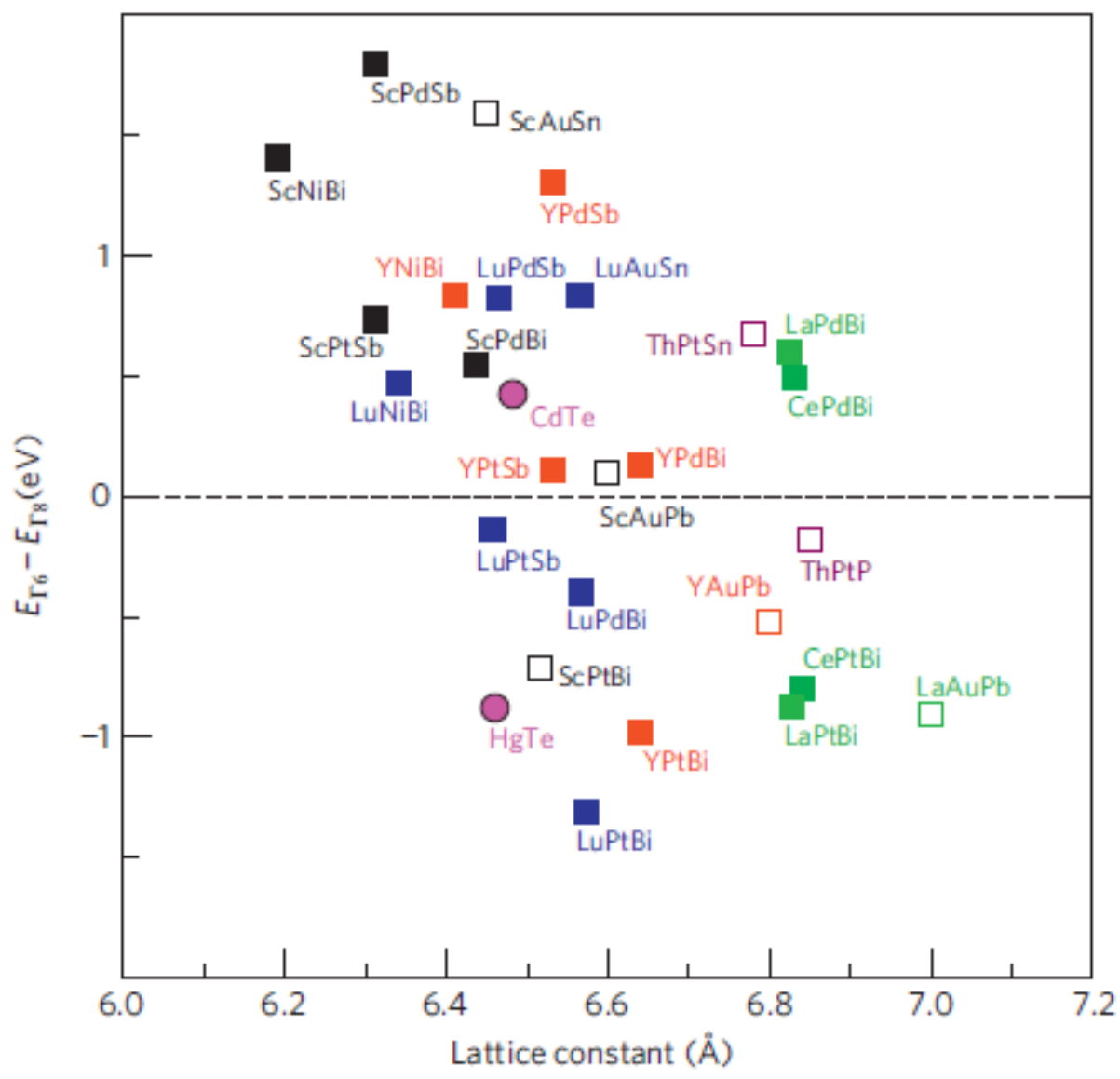

**Figure 35 Half-Heusler family of topological insulators. From Ref. [21].**

## 7.3. *TI and axion electrodynamics*

Similar to the Chern-Simons theory that Shoucheng investigated in the fractional quantum Hall effect, the new states QSH and 3D TI should be described by some topological field theories, which characterize their universal features. In early 2008, Xiao-Liang Qi, Taylor L. Hughes and Shoucheng developed the topological field theory of 3D TI [3]. Surprisingly, the field theory describing TI is related to electromagnetic duality and has appeared before in high energy physics as "axion electrodynamics" (For an introduction of the relation of TI and axion physics, see Frank Wilczek, Nature **458** 129 (2009)). The axion is a hypothetical particle in high energy physics with a particular coupling with the electromagnetic field $\theta E \cdot B$ [68]. In TI, $\theta$ is an angle determined by the material, with $\theta = 0, \pi$ corresponding to trivial insulator and TI. The effect of $\theta$ is a modification of the constituent equations:

$$\boldsymbol{D} = \boldsymbol{E} + 4\pi \boldsymbol{P} - \alpha \frac{\theta}{\pi} \boldsymbol{B}$$

$$\boldsymbol{H} = \boldsymbol{B} - 4\pi \boldsymbol{M} + \alpha \frac{\theta}{\pi} \boldsymbol{E}$$

The physical consequence of the axion coupling is a topological *magneto-electric* effect, which means an electric field induces magnetization, while a magnetic field induces charge polarization. The axion angle $\theta$ determines the amplitude of this effect. Different consequences of the topological magneto-electric effect have been proposed, including an image monopole induced by a charge near the surface [2] and a topological contribution to the Faraday and Kerr rotation of linear polarized light [3, 69]. The latter has been experimentally realized recently in 2016-2017 [36, 70, 71].

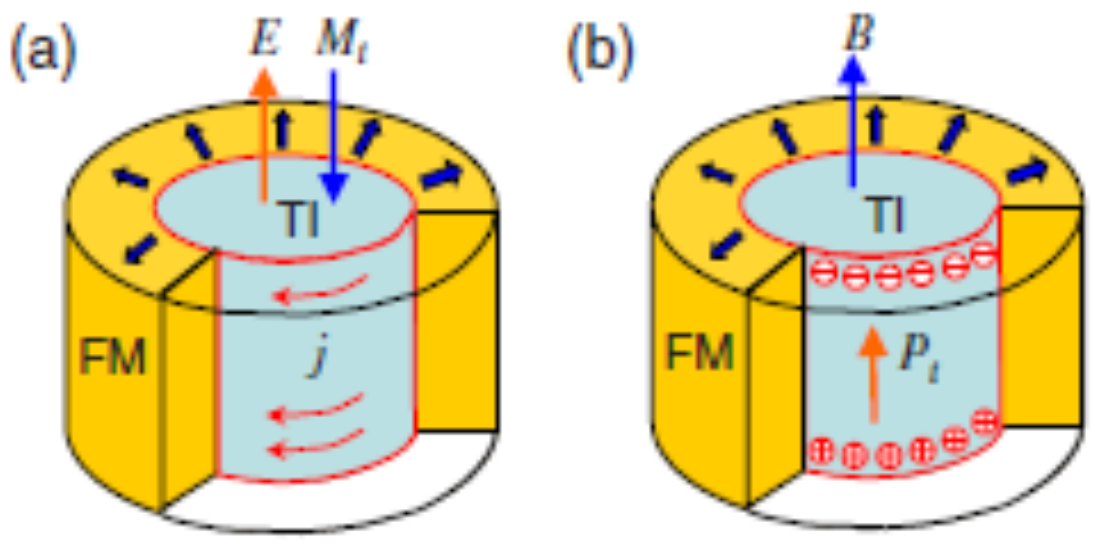

**Figure 36 Topological magneto-electric effect. From Ref. [3].**

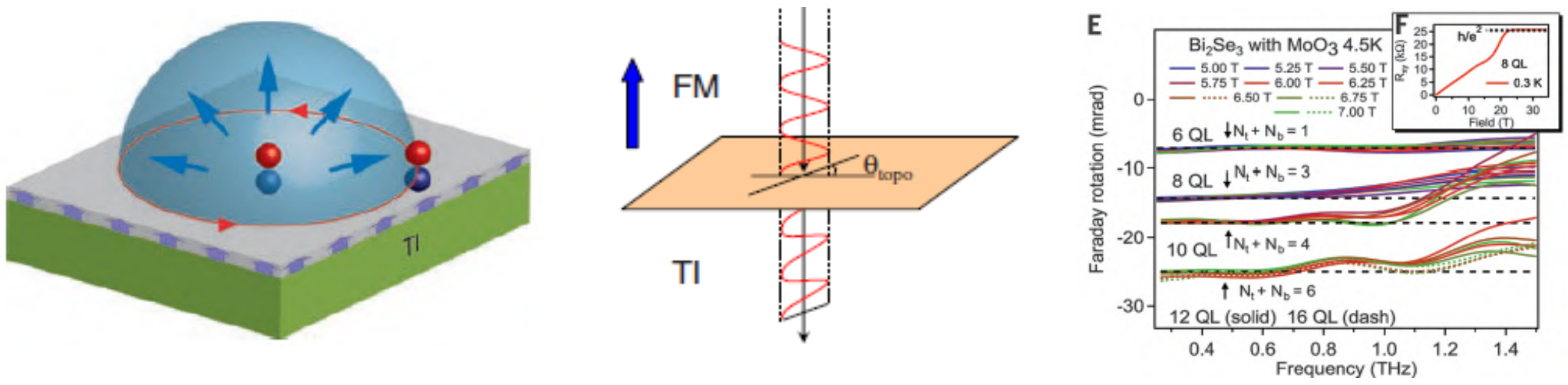

**Figure 37 (Left) Image magnetic monopole effect [2]. (Middle) Illustration of the topological Faraday rotation [3]. (Right) Faraday rotation experiment from Ref. [36].**

### 7.4. *Further generalization*

The topological field theory approach also provided a unified view to topological states in different dimensions and symmetry classes. The Qi-Hughes-Zhang paper proposed a family tree of topological insulators, in which the 4D QH state (determined by the second Chern number, a topological invariant) is identified as the "parent state" from which 3DTI and QSH are obtained by dimensional reduction. A different approach of classification was independently proposed by two other works [72, 73], which was based on non-interacting fermions but covered a larger family of topological states than [3]. Starting from these works, the family of topological states has greatly expanded, by considering new symmetry classes, interaction effects and gapless systems (topological semimetals). Shoucheng has made very broad contributions to many different aspects of this recent progress. In the next two sections we review two of these major directions, the quantum anomalous Hall (QAH) effect and topological superconductors (TSC).

## 8. Quantum Anomalous Hall Effect

### 8.1. *Quantum anomalous Hall effect*

The quantum anomalous Hall (QAH) effect refers to QHE in a band insulator without an external magnetic field. Topologically it belongs to the same phase as integer QH, but it has dispersing bands and no Landau levels. The first model for QAH was proposed by F. D. M. Haldane in 1988 [54]. Although in principle QAH does not require quantum spin Hall and spin-orbit coupling, it is only after the proposal of QSH and TI that a path to realize the QAH effect was found. In a 2006 paper Shoucheng collaborated on with X. L. Qi and Y. S. Wu about a quantum spin Hall model [74], the authors also proposed a new model of QAH. (This paper is possibly the origin of the name "quantum anomalous Hall effect".)

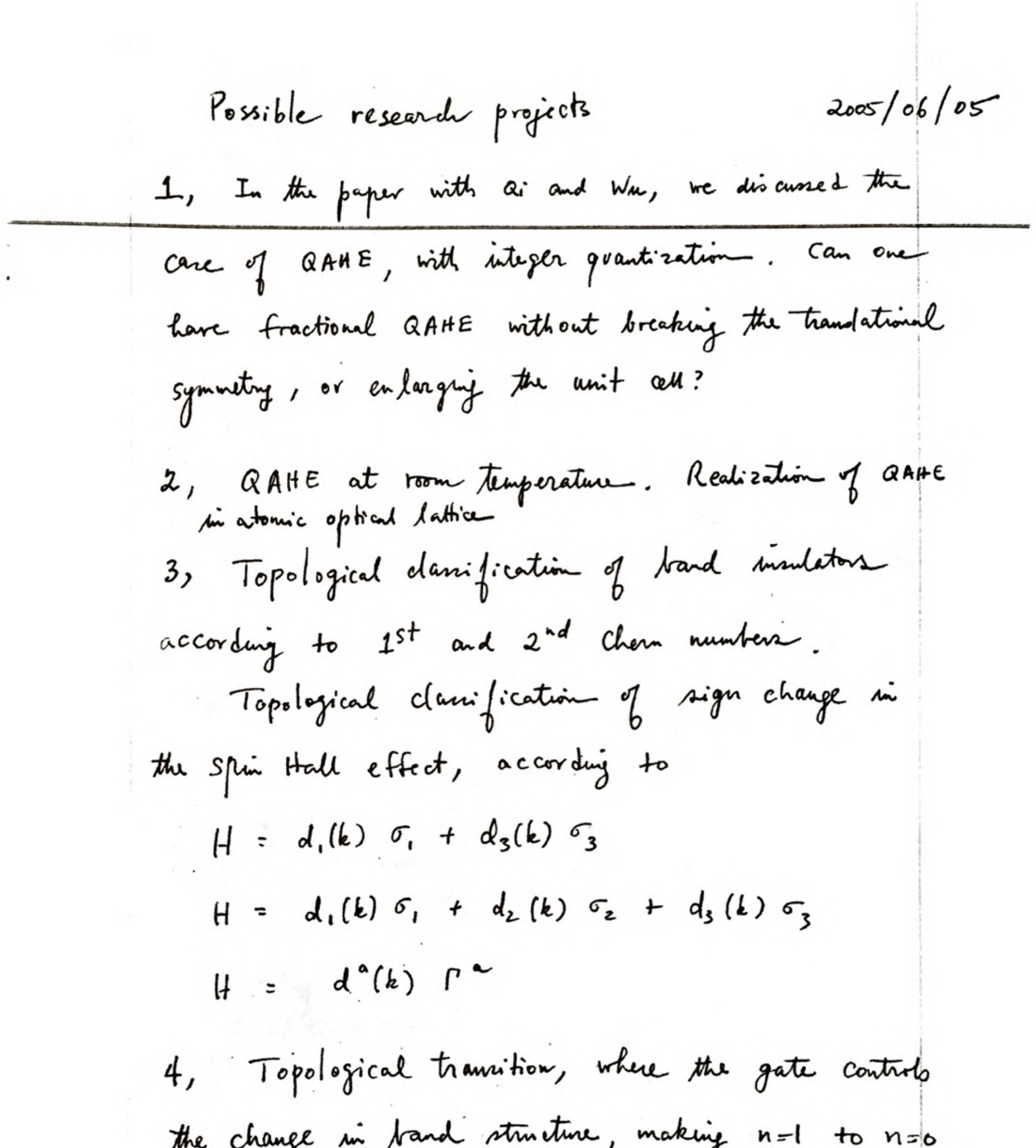

Possible research projects 2005/06/05

1, In the paper with Qi and Wu, we discussed the case of QAHE, with integer quantization. Can one have fractional QAHE without breaking the translational symmetry, or enlarging the unit cell?

2, QAHE at room temperature. Realization of QAHE in atomic optical lattice

3, Topological classification of band insulators according to $1^{st}$ and $2^{nd}$ Chern numbers.

Topological classification of sign change in the spin Hall effect, according to

$H = d_1(k)\,\sigma_1 + d_3(k)\,\sigma_3$

$H = d_1(k)\,\sigma_1 + d_2(k)\,\sigma_2 + d_3(k)\,\sigma_3$

$H = d^a(k)\,\Gamma^a$

4, Topological transition, where the gate controls the change in band structure, making $n=1$ to $n=0$ transition.

$n=0$ no edge channel

$n=1$ yes edge channel

**Figure 38 Shoucheng's note on a list of possible research projects related to QAHE in June 5, 2005. The first project is on fractional QAH effect, which became an active research field later in 2011. The second project of QAH effect at room temperature is still on one major goal of the field.**

QAH can be realized by making use of TI because the latter already has nontrivial Berry curvature that contributes to the Hall conductance. One just needs to break time-reversal in an appropriate way to avoid the cancellation.

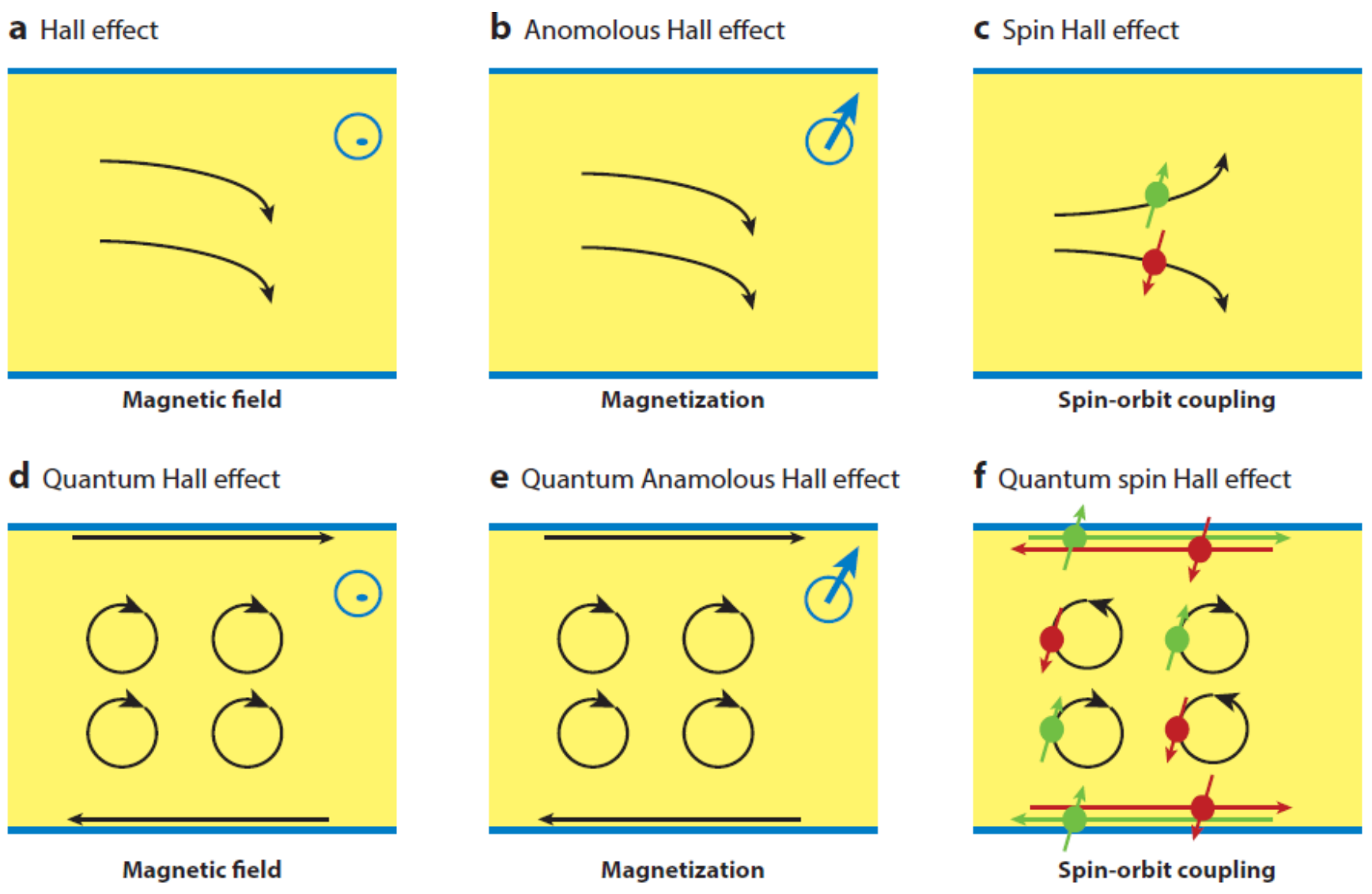

**Figure 39 Table of three Hall effects and their quantum versions. From Ref. [24].**

Based on further developments in TI, Shoucheng and collaborators made two more realistic proposals of QAH: one in Mn doped HgTe/CdTe quantum wells, making use of QSH [75]; the other in magnetically doped $Bi_2Se_3$ (or $Sb_2Te_3$, $Bi_2Te_3$) thin films, making use of the surface state of 3D TI [30]. The Hg(Mn)Te material is paramagnetic and therefore requires an external field, while the magnetically doped $Bi_2Se_3$ film was proposed to be ferromagnetic, which was experimentally realized in 2013 [18]. More recently, Shoucheng and his collaborators also identified a new family of intrinsic magnetic topological materials, $MnBi_2Te_4$, for the realization of the QAH effect [76]. The quantization of Hall resistance in this system has also been experimentally observed [77].

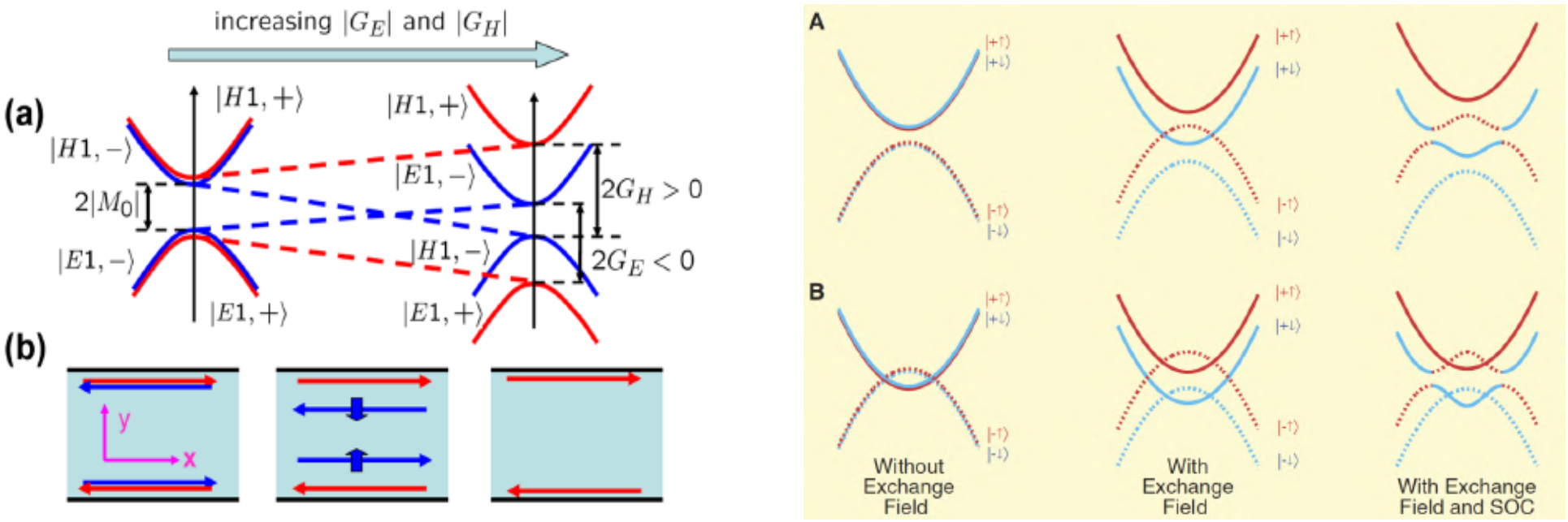

**Figure 40 Band inversion mechanism for QAH effect in Mn doped HgTe (left) and magnetically doped $Bi_2$ $Se_3$ film. From Ref. [12] and Ref. [30].**

The QAH in magnetically doped $Bi_2Se_3$ family TI was first realized in Cr-doped $(Bi,Sb)_2Te_3$ thin films in Qikun Xue's group in Tsinghua University [18]. Shoucheng played an important role in this collaboration. Since then, the QAH state has been firmly verified in more and more studies. Many unique features of the QAH state, such as zero Hall plateau [78], universal scaling behavior at the plateau transition [79], QAH state with higher plateaus [80], and anomalous edge transport due to coexisting chiral and helical modes [81], were studied by Shoucheng's group. These works guided the rapid development of this field and many of the theoretical predictions have been demonstrated in experiments [82-86].

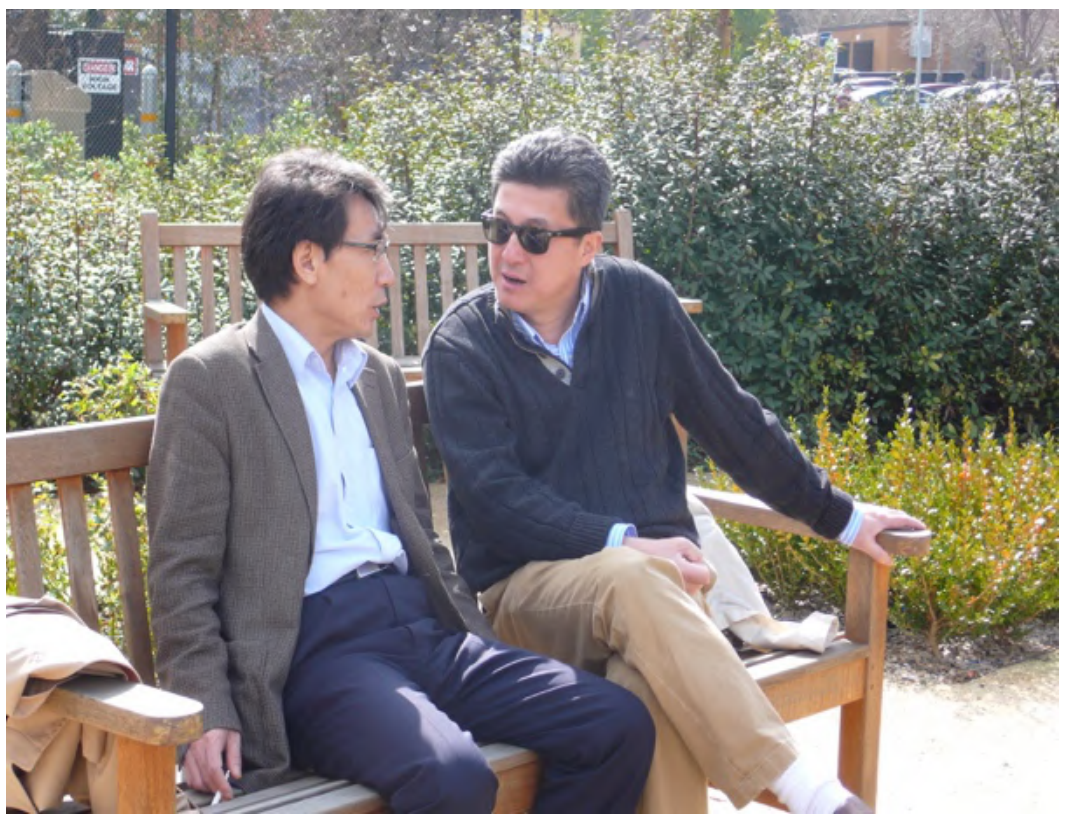

**Figure 41 Qikun Xue and Shoucheng in a discussion at Stanford campus, 2014**

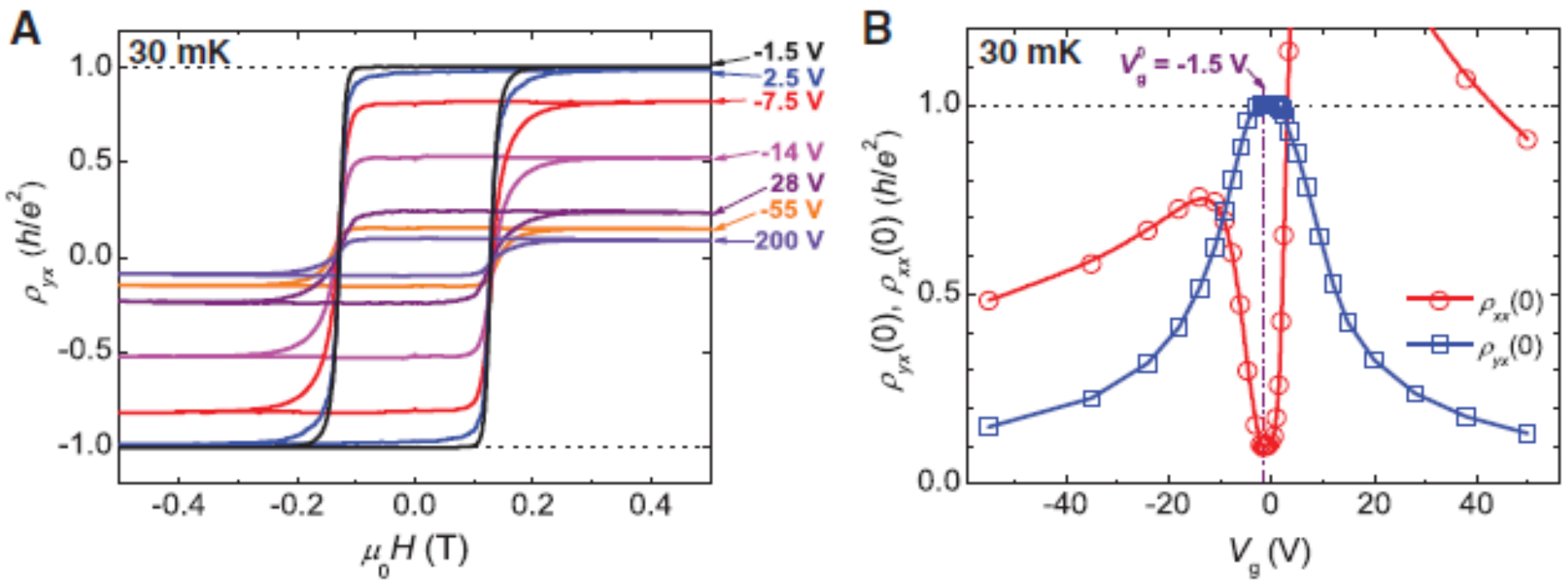

**Figure 42 (Left) Hall resistivity vs magnetic field. (Right) Gate voltage dependence of Hall resistivity and longitudinal resistivity. From Ref. [18].**

# 9. Topological Superconductors and More

## 9.1. *Topological superconductors*

The BCS mean-field theory of superconductivity describes quasiparticles in a similar way as electrons in band insulators. When the superconductor has a full gap, the classification is very similar to that of insulators, which suggested the concept of topological superconductors (TSC). (This is analogous to how Ettore Majorana generalized the Dirac equation to his Majorana equation in 1937.) Historically, the first topologically nontrivial superconductors discovered were 2D p+ip superconductors [87] and 1D p-wave superconductors [88].

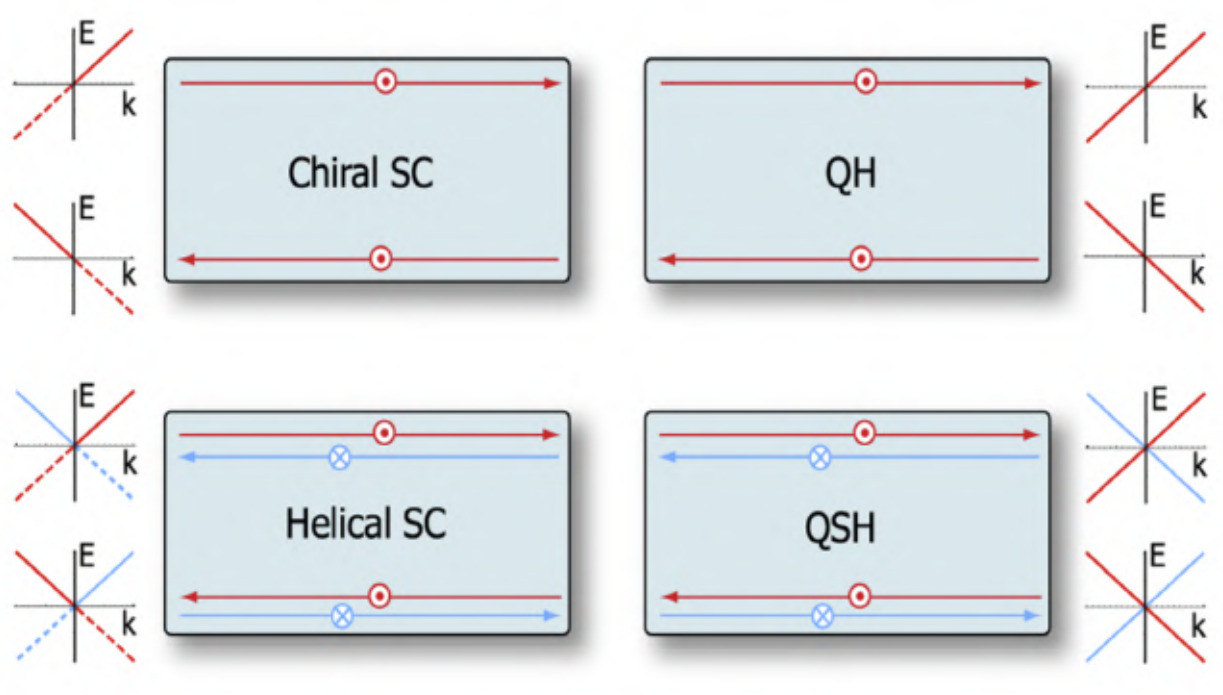

**Figure 43 Comparison between 2D time-reversal invariant TSC and QSH, from Ref. [23]**

Following the discovery of time reversal invariant (TRI) topological insulators, new TSC with time reversal symmetry were proposed in 2008-2009, by independent works from Shoucheng's group and other groups [23, 72, 73]

Interestingly, the B phase of $^3$He superfluid, known from many years ago, was proposed as a 3D TRI topological superfluid.

### 9.2. *2D chiral TSC from QAH*

The 2D p+ip TSC (also known as chiral TSC) is an interesting system for quantum computation, because of non-Abelian statistics of vortices with Majorana zero modes. A candidate material for chiral TSC is $Sr_2RuO_4$, but it has not been confirmed. In 2010, Shoucheng's group proposed a new mechanism for realizing chiral TSC, by making use of QAH effect [1]. The proposal is that in general a quantum Hall phase transition will be broadened into a TSC phase when superconducting proximity effect is introduced. If realized, this proposal also enables new way to probe TSC through the QAH edge states. Shoucheng's group proposed that the chiral Majorana fermions yield a half quantum Hall plateau in a hybrid device of QAH and TSC [32, 89]. Experimental evidence of half quantum Hall plateau was reported in 2017 [90], although the interpretation remains controversial [91, 92]. Based on this proposal, in 2018 Shoucheng's group further proposed the possibility of realizing non-Abelian quantum gates with chiral Majorana fermions [11].

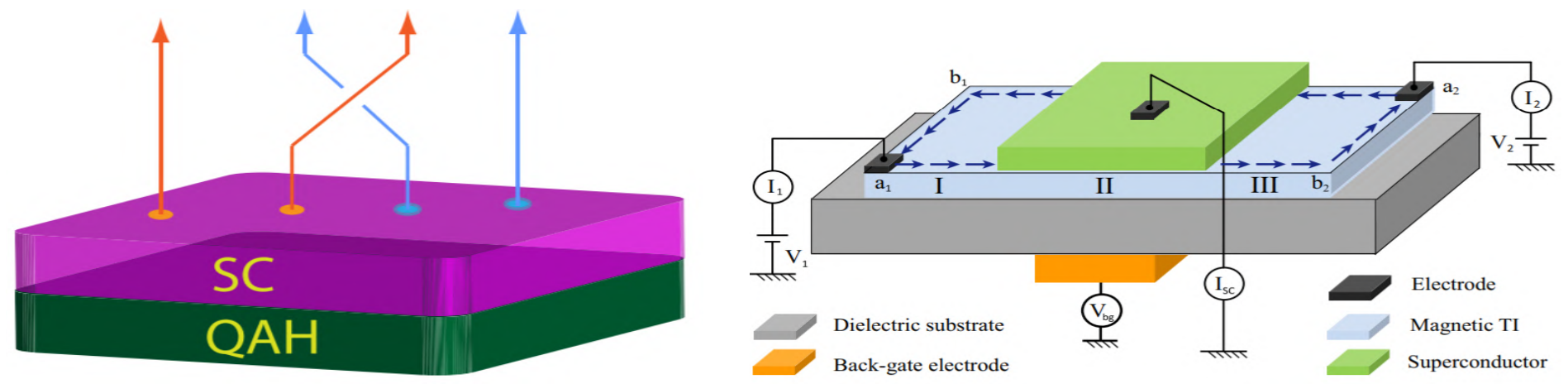

**Figure 44 Shoucheng's proposals for 2D chiral TSC [1] and device for detecting chiral Majorana fermion [32].**

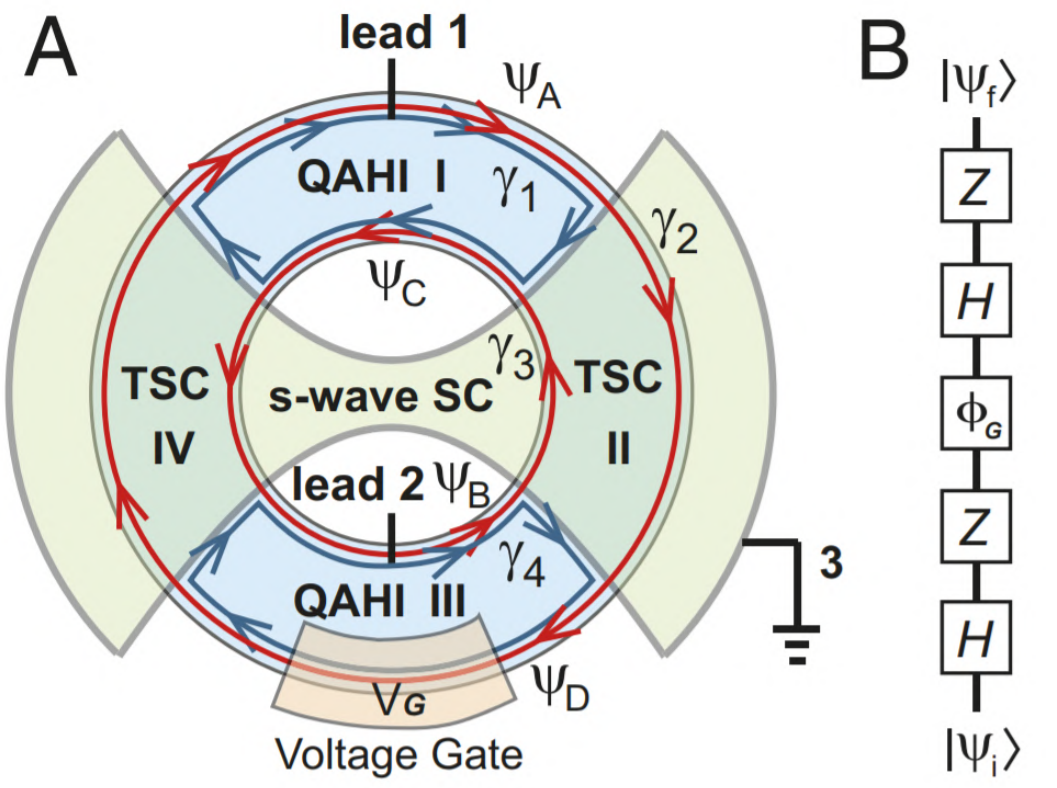

**Figure 45 A single qubit quantum gate of chiral Majorana fermions [11].**

## 10. Other areas of condensed matter physics

Besides the directions discussed so far, Shoucheng also conducted research in a wide variety of other topics in physics, such as topological semimetals, interacting topological states, Fe-pnictide superconductors, the role of symmetry in quantum Monte Carlo method, etc. For example, in Ref. [42], Shoucheng and Congjun Wu proposed a new idea in dealing with the long-standing sign problem of quantum Monte Carlo method. The latter is a widely used numerical method in quantum many-body physics and statistical mechanics, which allows an efficient sampling of a probability distribution. A generic quantum problem cannot be reduced to evaluating the expectation value of simple physical quantity in a probability distribution, due to the complex phase of the wavefunction. This is known as the sign problem. Ref. [42] pointed out how the sign problem can be avoided in some systems by making use of discrete symmetries such as time reversal symmetry and particle-hole symmetry. This method has been applied to a wide variety of problems (see e.g. Z. C. Wei, Congjun Wu, Yi Li, Shiwei Zhang, and T. Xiang Phys. Rev. Lett. 116, 250601) As another example, at the early stage of the discovery of Fe-pnictide high-temperature superconductors, in Ref. [40], Shoucheng and collaborators proposed the first minimal model for this system, which has influenced a lot of later works in this direction.

Shoucheng has also collaborated with high-energy physicists on interdisciplinary problems related to topological physics. For example, in Ref. [38] in collaboration with Edward Witten and Xiao-Liang Qi, they related the physics of three-dimensional time-reversal invariant topological superconductor to anomaly inflow and axion theory studied in high energy physics. In Ref. [39], in collaboration with Biao Lian, Cumrun Vafa and Farzan Vafa, they showed that the low-energy field theory of topological nodal line superconductors is equivalent to Wilson loops in the Chern-Simons theory, which relates the linking invariants of nodal lines to thermal magnetoelectric effect in a superconductor. We end this section with a selection of figures from Shoucheng's papers (Figure 46)), which provide a glance into his diverse research interests.

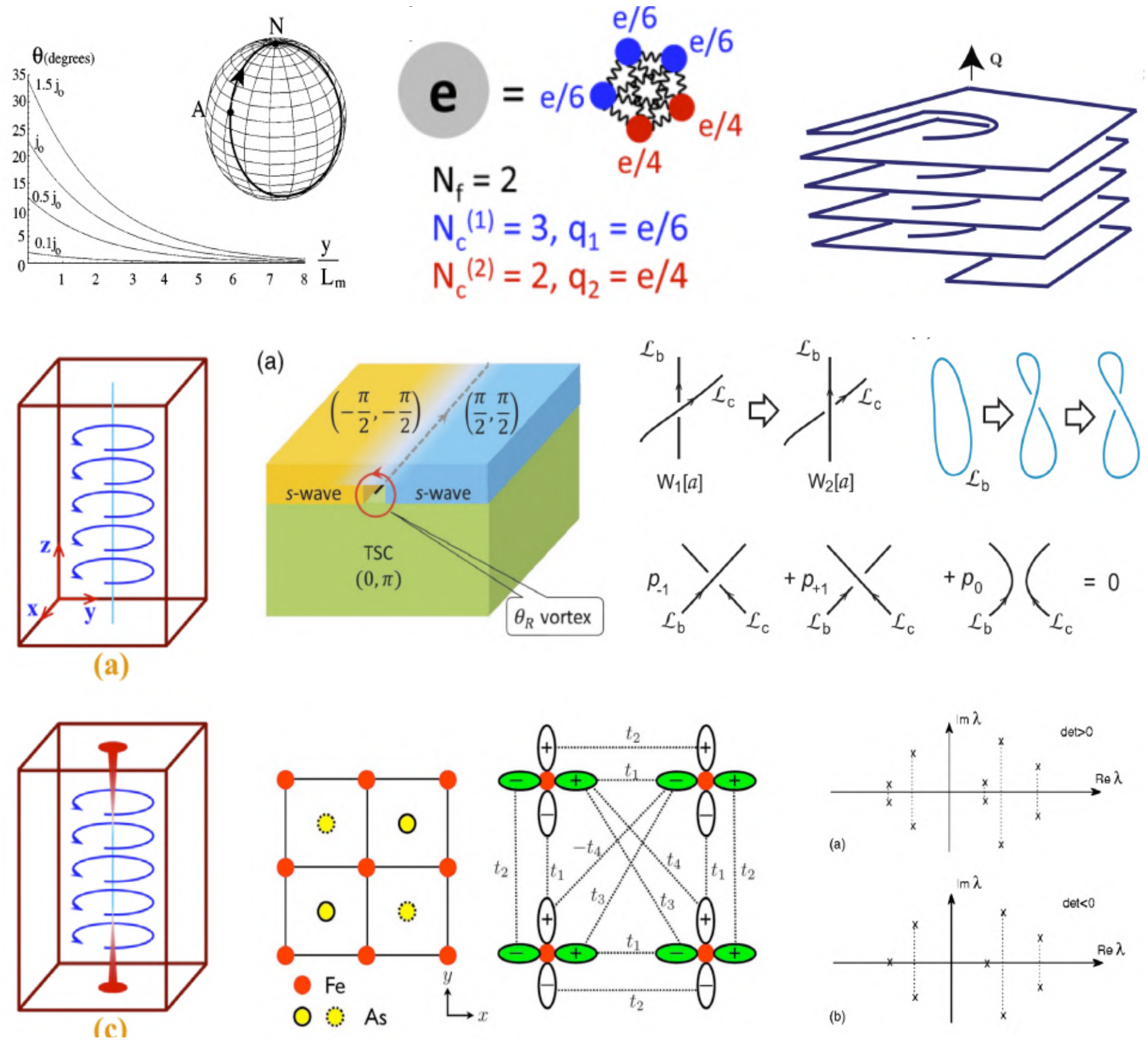

**Figure 46 Snapshots of Shoucheng's works on (starting from top left) giant-magnetoresistive materials [22], fractional TI [26, 27], Weyl semimetals and axion strings [34], Fe-pnictide topological vortex [37], axion theory of TSC [38], nodal semimetals and Chern-Simons theory [39], Fe-pnictide effective model [40], time reversal and quantum Monte Carlo [42].**

## 11. Exploration beyond physics

Shoucheng always believed that science should have no disciplinary borders. He often told his students to spend 80% of their time on their research focus, and 20% time exploring broader directions and thinking about big questions. In recent years, Shoucheng investigated many directions beyond physics such as artificial intelligence (AI), blockchain, distributed computing, bioinformatics, and more. Just like the previous years, Shoucheng kept a collection of well-organized handwritten notes in 2018. In Appendx A, we display a few pages of this precious record, from which we can see how many different directions he was exploring actively until the last days of his life.

Shoucheng's ideas often uncovered new connections between fundamental physics principles and various other fields. In 2018, Shoucheng and collaborators used AI to study chemical compounds [20] (Q. Zhou et. al., *PNAS* **115**, 6411 (2018)). They developed an "Atom2Vec" algorithm and showed that the AI did not only rediscover the periodic table but was also able to predict new chemical compounds from the properties of elements it discovered.

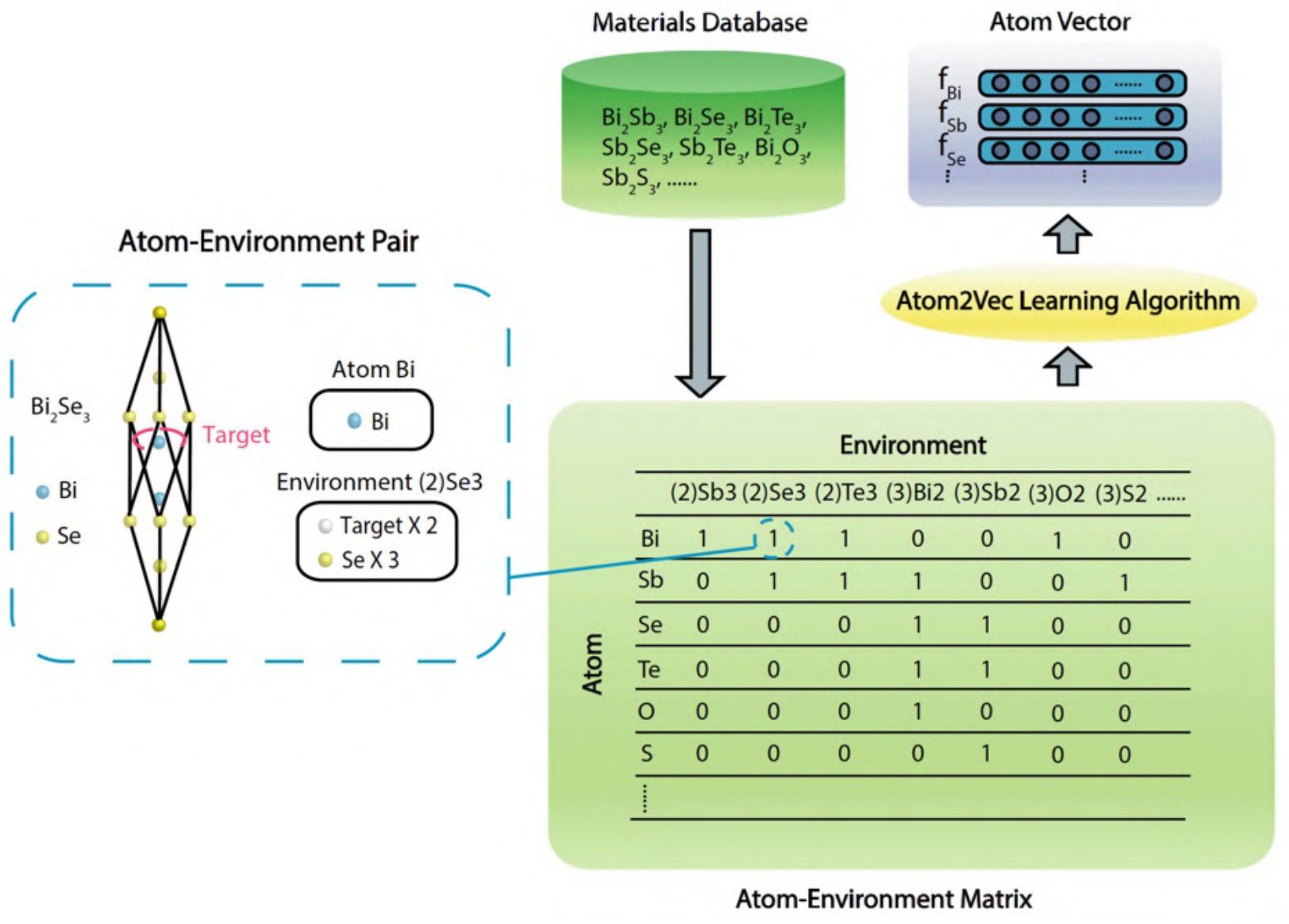

**Figure 47 Illustration of the Atom2Vec algorithm. From Ref. [20].**

Another area of Shoucheng's recent interest was blockchain technology. Blockchain (and its generalizations) is a mathematical mechanism to reach consensus on a record without requiring any centralized agency. In analogy with the concept of information entropy, Shoucheng emphasized that consensus has an intrinsic value, and envisioned that "humanity is now reaching a new era where trust and consensus is built upon math," a viewpoint that he summarized in the motto "in math we trust." Shoucheng believed that new consensus mechanisms would help to build a fairer, more efficient, and more diverse society.

Shoucheng was also interested in many other areas related to how mathematics and algorithms could improve human society. An example is the application of secure multi-party computation in bioinformatics and medical sciences. Shoucheng's son Brian Zhang suggested a common theme behind the different areas of Shoucheng's latest interests, which is "algorithms for the future."

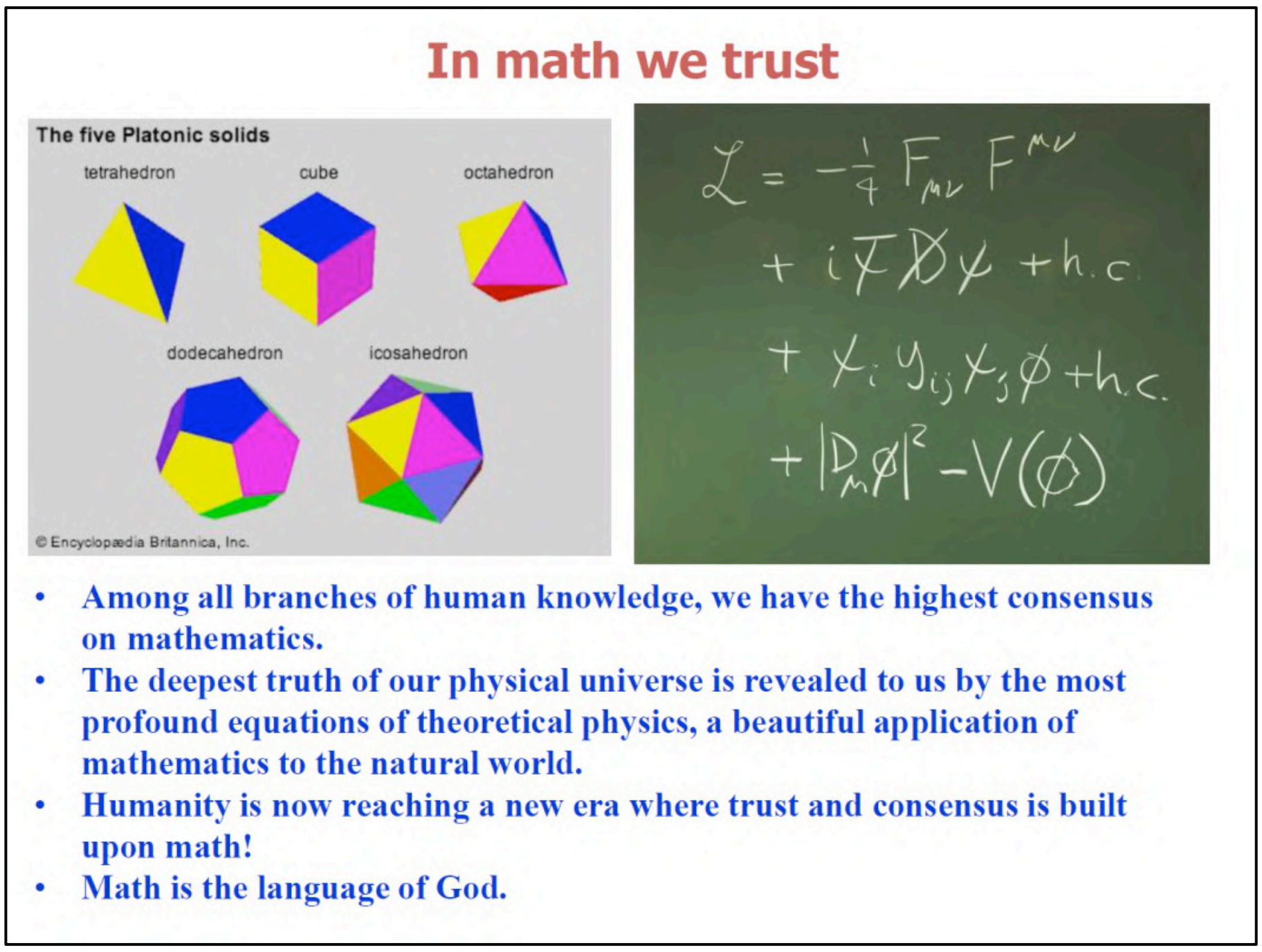

**Figure 48 A slide from Shoucheng's talk "In Math We Trust – Foundations of Crypto-economic Science" in Sep. 2018.**

## 12. To see a world in a grain of sand

There are many aspects of Shoucheng's legendary scientific life that we are not able to cover in our brief overview. Despite his abrupt departure, his groundbreaking scientific works; his dreams and vision for science, life, and humanity; and his inspiration to all of us will forever be preserved as an eternal monument to him. We would like to end this article by his favorite quote from William Blake's poem:

***To see a World in a Grain of Sand***
***And a Heaven in a Wild Flower***
***Hold Infinity in the palm of your hand***
***And Eternity in an hour***

**Acknowledgement.** We would like to thank the Zhang family members Barbara, Stephanie, Brian, and Ruth for sharing with us the precious photos and notes of Shoucheng, and for providing a lot of helpful suggestions. We would like to thank Yaroslaw Bazaliy, B. Andrei Bernevig, Leonid Pryadko, and Congjun Wu for helpful suggestions. We would like to thank all authors of this memorial volume, and all other speakers and participants of the Shoucheng Zhang Memorial Workshop.

## Appendix A. Shoucheng's notes in 2018

In this appendix, we include the snapshot of a few pages from Shoucheng's handwritten notes in 2018 on various topics. These are a small part of the many topics that he was learning about and/or brainstorming about. Starting from the first page, it covers the following topics:

1. CRISPR (clustered regularly interspaced short palindromic repeats, CAS9), which is a simple powerful tool for editing genomes.
2. Vector clock and causality, which is a mathematical data structure for determining the partial ordering of events (i.e., determining time) in distributed systems.
3. CAS12 based molecular diagnosis, which is a method similar to CAS9 for editing genomes.
4. DNA finger printing, the technique for identifying an individual from a sample of DNA by looking at unique patterns in their DNA.
5. DAG (directed acyclic graph), which is a mathematical concept in graph theory. Shoucheng was interested in its application in the generalization of the block chain.
6. A simple review of ZKP (zero knowledge proofs), which is a method by which one can prove to another person that he/she knows some knowledge without conveying any information apart from the fact that they know the knowledge.

7. Maxwell's demon. A thought experiment proposed by James C. Maxwell about a hypothetical way of violating the second law of thermodynamics. The note is about one way of understanding the resolution of the paradox.

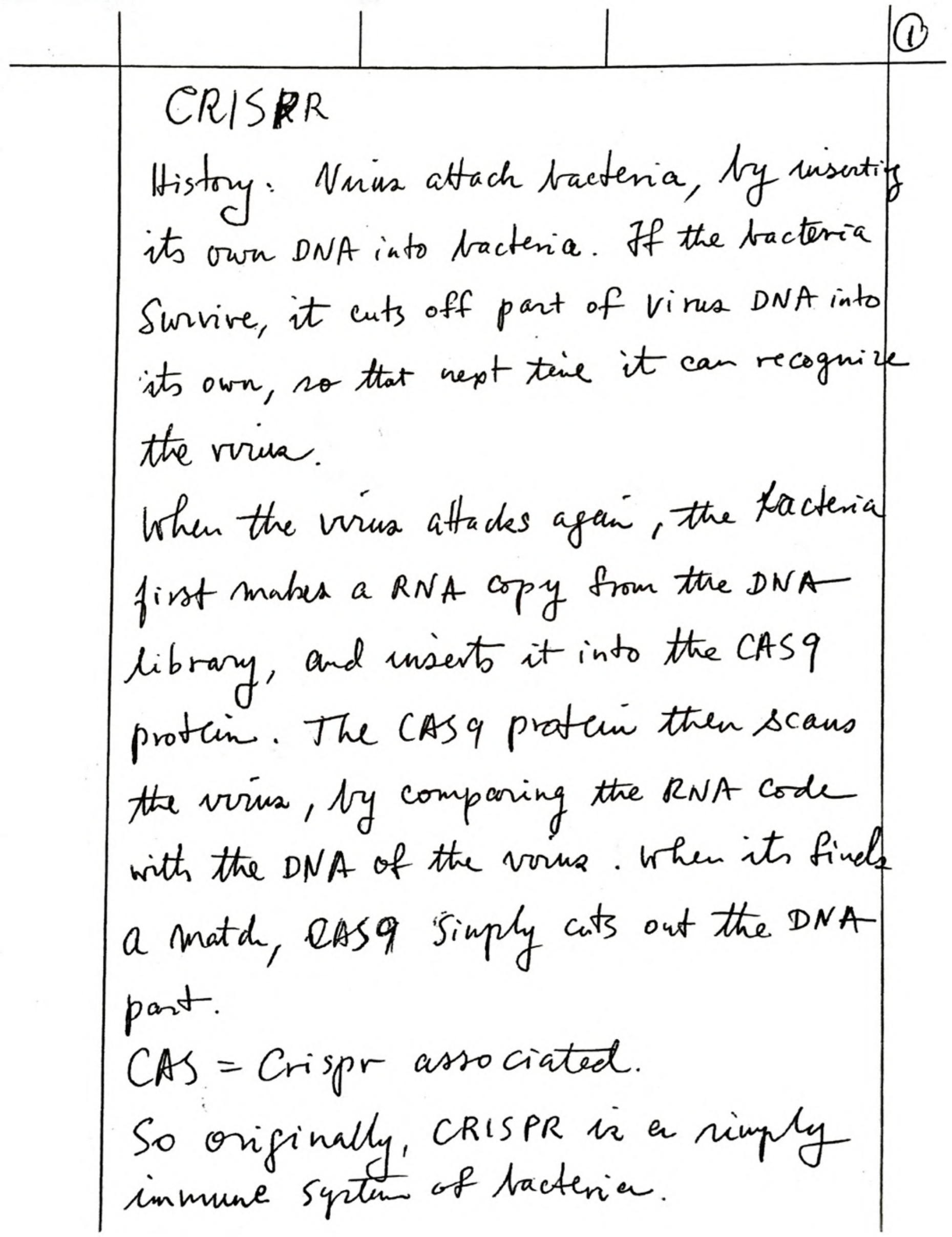
①

CRISPR

History: Virus attach bacteria, by inserting its own DNA into bacteria. If the bacteria survive, it cuts off part of virus DNA into its own, so that next time it can recognize the virus.

When the virus attacks again, the bacteria first makes a RNA copy from the DNA library, and inserts it into the CAS9 protein. The CAS9 protein then scans the virus, by comparing the RNA code with the DNA of the virus. When its finds a match, CAS9 simply cuts out the DNA part.

CAS = Crispr associated.

So originally, CRISPR is a simply immune system of bacteria.

**Figure 49 Shoucheng's note on CRISPER (clusters of regularly interspaced short palindromic repeats)**

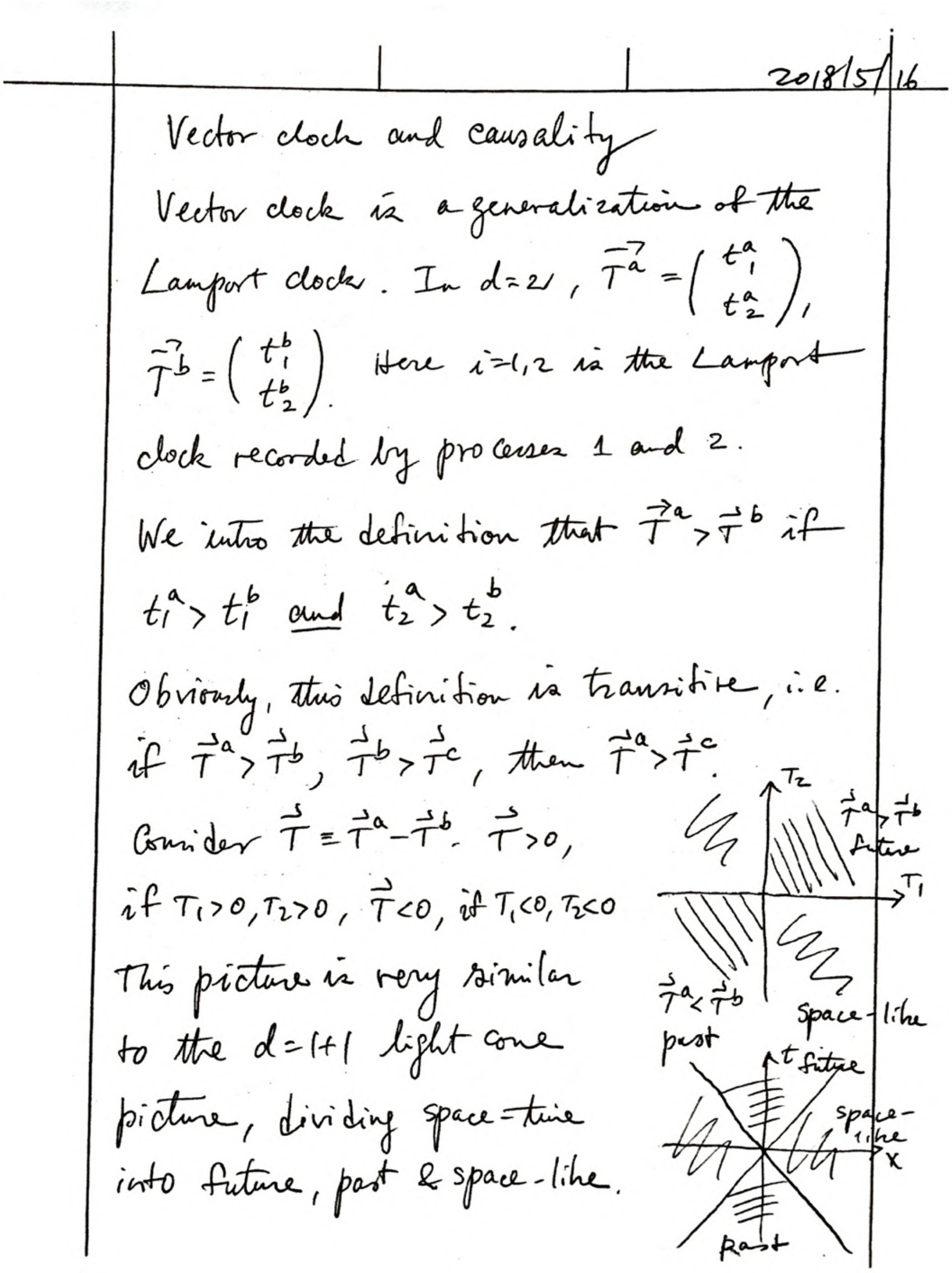

2018/5/16

Vector clock and causality

Vector clock is a generalization of the Lamport clock. In $d=2$, $\vec{T}^a = \begin{pmatrix} t_1^a \\ t_2^a \end{pmatrix}$, $\vec{T}^b = \begin{pmatrix} t_1^b \\ t_2^b \end{pmatrix}$. Here $i=1,2$ is the Lamport clock recorded by processes 1 and 2.

We intro the definition that $\vec{T}^a > \vec{T}^b$ if $t_1^a > t_1^b$ <u>and</u> $t_2^a > t_2^b$.

Obviously, this definition is transitive, i.e. if $\vec{T}^a > \vec{T}^b$, $\vec{T}^b > \vec{T}^c$, then $\vec{T}^a > \vec{T}^c$.

Consider $\vec{T} = \vec{T}^a - \vec{T}^b$. $\vec{T} > 0$, if $T_1 > 0, T_2 > 0$, $\vec{T} < 0$, if $T_1 < 0, T_2 < 0$

This picture is very similar to the $d=1+1$ light cone picture, dividing space-time into future, past & space-like.

**Figure 50 Shoucheng's note on vector clock and causality.**

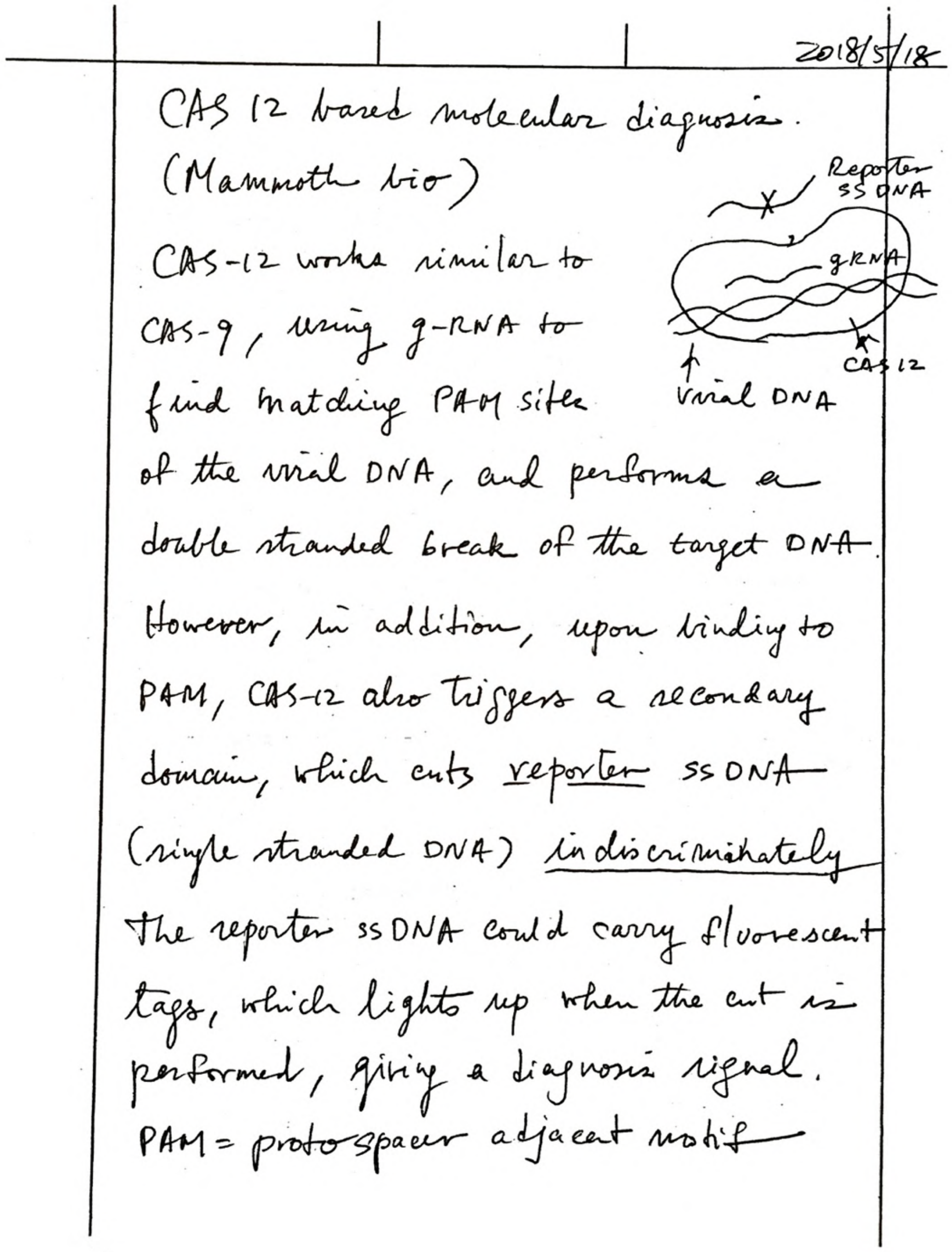
2018/5/18

CAS 12 based molecular diagnosis.
(Mammoth bio)

CAS-12 works similar to CAS-9, using g-RNA to find matching PAM sites of the viral DNA, and performs a double stranded break of the target DNA.
However, in addition, upon binding to PAM, CAS-12 also triggers a secondary domain, which cuts reporter ssDNA (single stranded DNA) indiscriminately
The reporter ssDNA could carry fluorescent tags, which lights up when the cut is performed, giving a diagnosis signal.
PAM = protospacer adjacent motif

**Figure 51 Shoucheng's note on CAS12 based molecular diagnosis.**

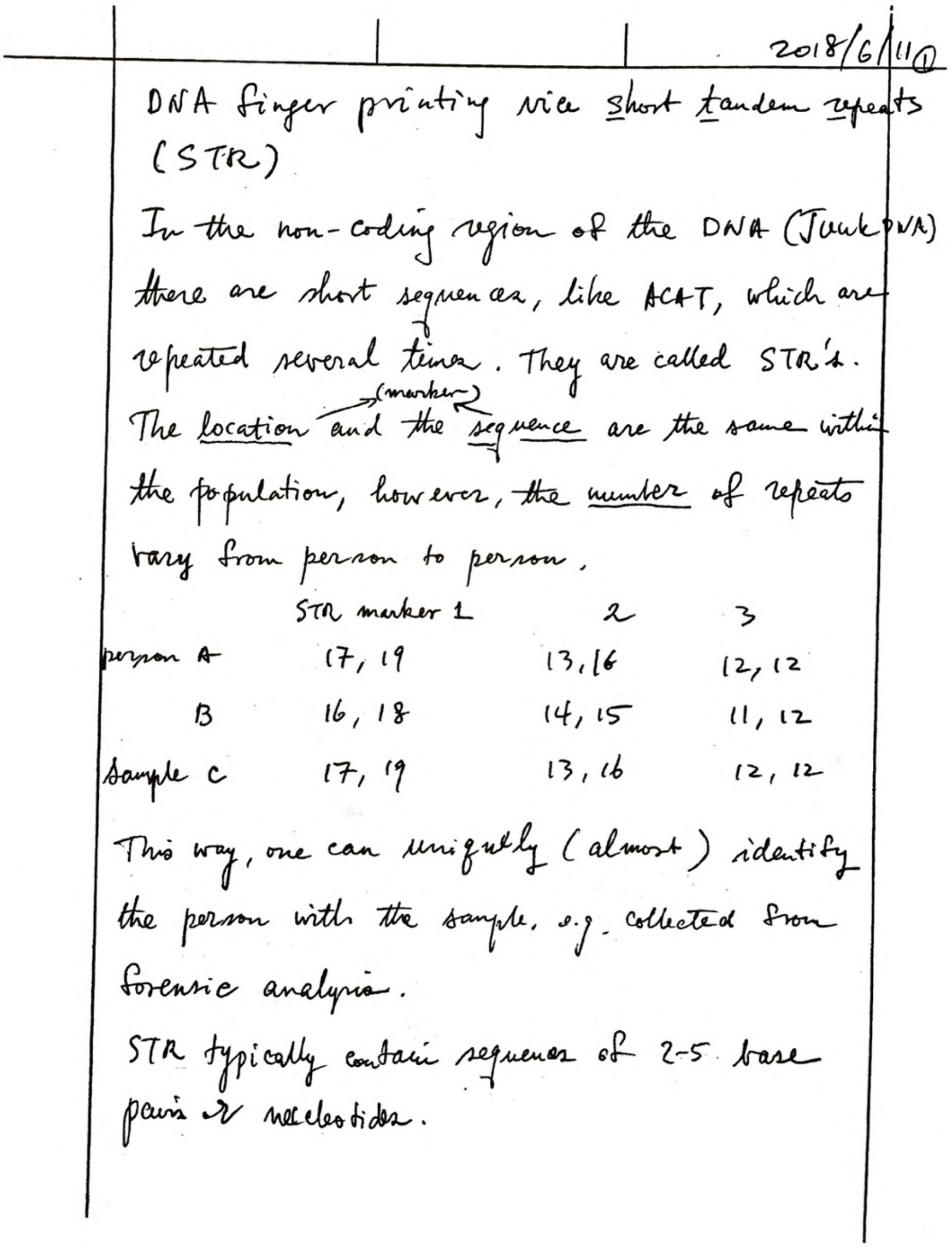

2018/6/11 ①

DNA finger printing via short tandem repeats (STR)

In the non-coding region of the DNA (Junk DNA) there are short sequences, like ACAT, which are repeated several times. They are called STR's.

The location and the sequence (marker) are the same within the population, however, the number of repeats vary from person to person.

| | STR marker 1 | 2 | 3 |
|---|---|---|---|
| person A | 17, 19 | 13, 16 | 12, 12 |
| B | 16, 18 | 14, 15 | 11, 12 |
| Sample C | 17, 19 | 13, 16 | 12, 12 |

This way, one can uniquely (almost) identify the person with the sample, e.g. collected from forensic analysis.

STR typically contain sequences of 2-5 base pairs or nucleotides.

**Figure 52 Shoucheng's note on DNA finger printing.**

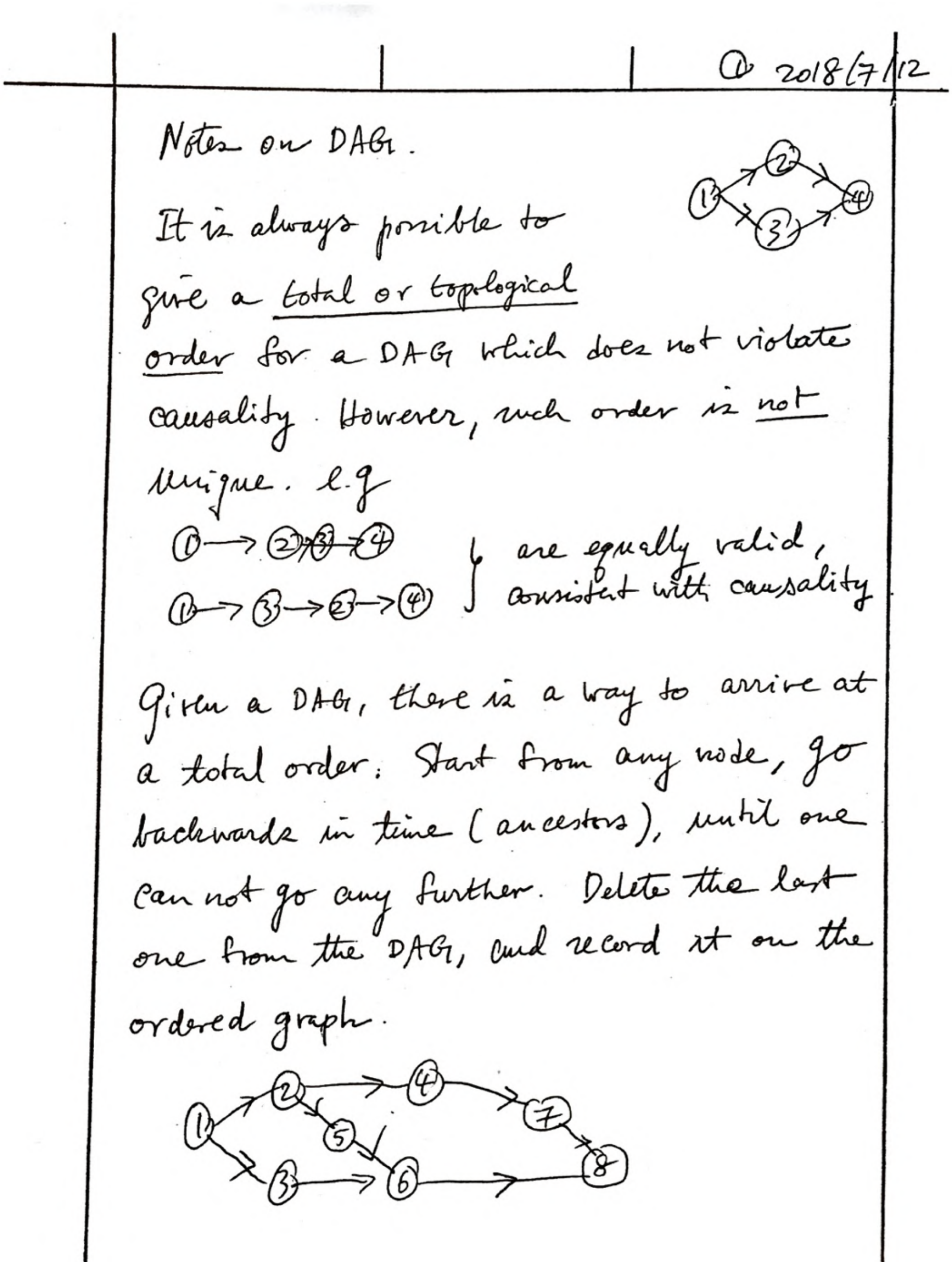

① 2018/7/12

Notes on DAG.

It is always possible to give a total or topological order for a DAG which does not violate causality. However, such order is not unique. e.g.

①→②→③→④
①→③→②→④ } are equally valid, consistent with causality

Given a DAG, there is a way to arrive at a total order: Start from any node, go backwards in time (ancestors), until one can not go any further. Delete the last one from the DAG, and record it on the ordered graph.

**Figure 53 Shoucheng's note on DAG (directed acyclic graph) and its application in block chain.**

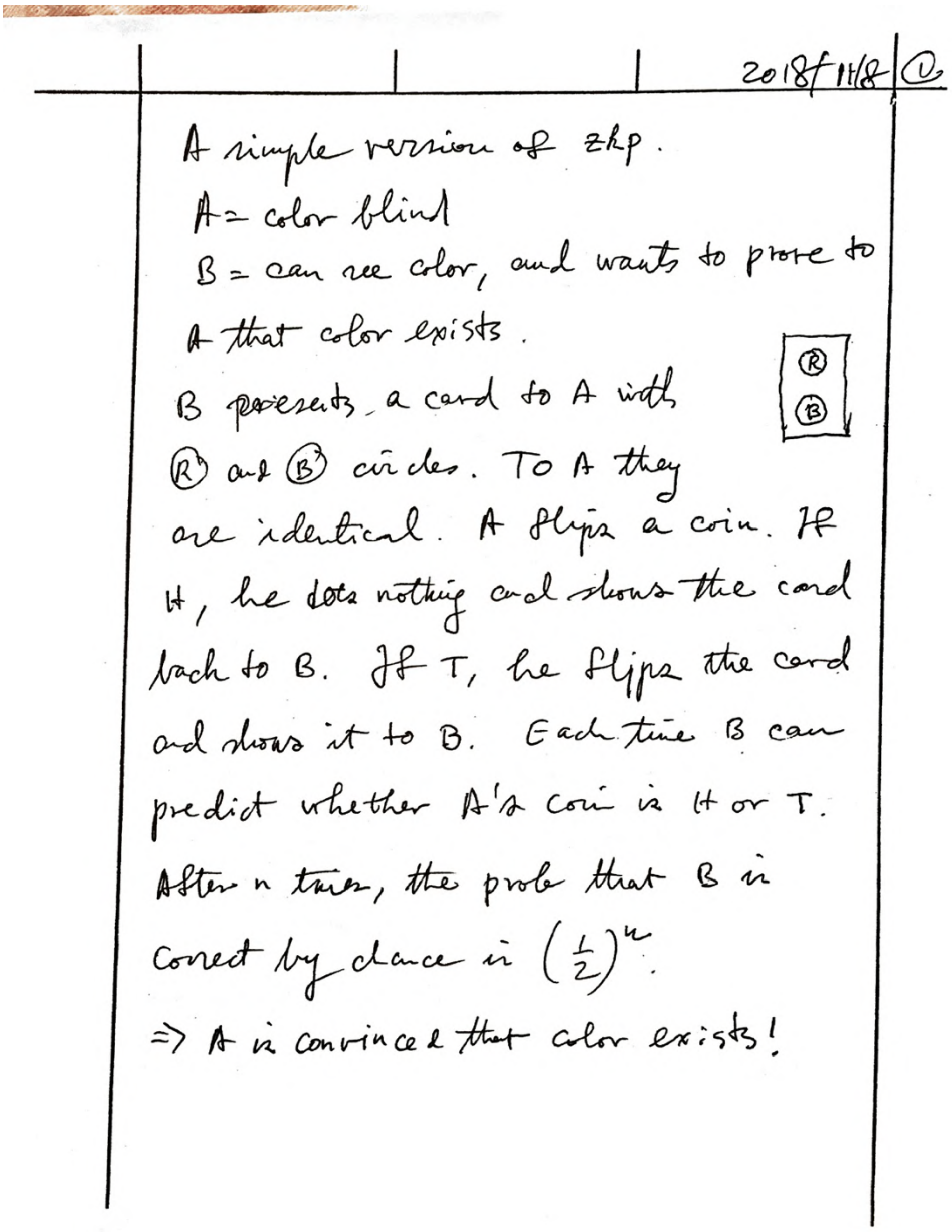
2018/11/8 ①

A simple version of ZKP.

A = color blind

B = can see color, and wants to prove to A that color exists.

B presents a card to A with (R) and (B) circles. To A they are identical. A flips a coin. If H, he does nothing and shows the card back to B. If T, he flips the card and shows it to B. Each time B can predict whether A's coin is H or T. After n tries, the prob that B is correct by chance is $\left(\frac{1}{2}\right)^n$.

=> A is convinced that color exists!

**Figure 54 Shoucheng's note on zero knowledge proofs.**

(1)

Maxwell's demon

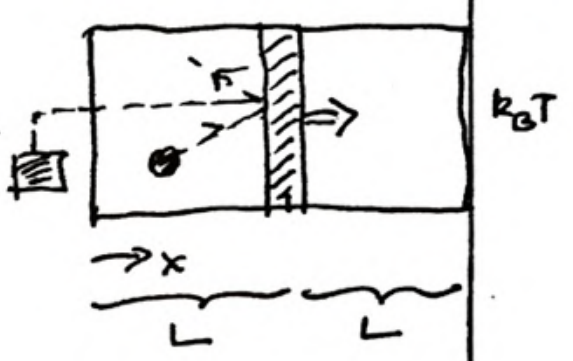

We consider a container with only one particle, in contact with a thermal bath at T. The demon is trying to extract some useful work out of the random motion of the particle.

① The demon inserts a wall at the middle.

② If the demon "knows" that the particle is on the left, he attaches a weight on the left side. If he knows that the particle is on the right, he attaches the weight on the right.

③ The wall can move without friction. and will expand isothermally, doing work

$$W = \int A p(x)\, dx = \int A \frac{k_B T}{A x} dx = k_B T \ln x \Big|_L^{2L}$$

$$= k_B T \ln 2$$

**Figure 55 Shoucheng's note on Maxwell's demon.**